\documentstyle[preprint,aps,prl,eqsecnum,epsfig]{revtex}
\newcommand{\beq}{\begin{equation}}
\newcommand{\eeq}{\end{equation}}
\newcommand{\bea}{\begin{eqnarray}}
\newcommand{\eea}{\end{eqnarray}}

\newcommand\refpar[1]{(\ref{#1})}

\newcommand{\U}{\mbox{\boldmath $U$}}
\newcommand{\pu}{\mbox{\boldmath $u$}}
\newcommand{\V}{\mbox{\boldmath $V$}}
\newcommand{\pv}{\mbox{\boldmath $v$}}
\newcommand{\cphase}{c_{\phi}}

\begin{document}
\draft
%
\title{Scattering of dislocated wavefronts by vertical vorticity and the
Aharonov-Bohm effect II : Dispersive waves}
\author{{\sc Christophe Coste}$^a$,  and
{\sc Fernando Lund}$^b$ \\[1em]
$^a$ Laboratoire de Physique, ENS Lyon\\
46, All\'ee d'Italie 69364 Lyon Cedex 07, France\\
 $^{b}$Departamento de F\'\i sica, Facultad de
Ciencias F\'\i sicas y Matem\'aticas \\
 Universidad de Chile, Casilla 487-3,
Santiago, Chile}

\maketitle

\begin{abstract}
Previous results on the scattering of surface waves by
vertical vorticity on shallow water are generalized to
the case of dispersive water waves.  Dispersion effects are treated
perturbatively around the shallow water limit, to first order in the ratio
of
depth to wavelength.  The dislocation of the incident wavefront, analogous
to the Aharonov-Bohm effect, is still observed.  At short wavelengths
the scattering is
qualitatively similar to the nondispersive case.  At moderate wavelengths,
however, there are two markedly different scattering regimes according to
wether the capillary length is smaller or larger than $\sqrt{3}$ times
depth. The dislocation is
characterized by a parameter that depends both on phase and group
velocity.
The validity range of the calculation is the same as
in the shallow water case: wavelengths small compared to vortex radius, and low
Mach number. The implications of these limitations  are carefully
considered.
\end{abstract}


\section{Introduction}
\label{sec:introduction}
In a preceding paper \cite{coumlu}, hereafter refered to as I, we studied
the scattering of surface waves by a stationary vertical vortex in the
long wavelength approximation: surface tension was neglected and the
fluid depth was supposed to be small compared to wavelength. This is also
called the shallow water approximation.
There
were two motivations for the study of shallow water waves scattering.
First, they
are non dispersive waves, like acoustic waves in fluids, and it was
plausible
that a generalization of calculations for sound scattering by vorticity
\cite{umlu} was feasible. Secondly, it was a first attempt towards a
quantitative
confirmation of the heuristic approach of Berry {\it et al.}
\cite{berryetal}. The aim of this paper is to go beyond this approximation.

In actual experimental situations \cite{vivancomelo} the shallow water
limit is
hard to obtain and, if a quantitative comparison with experiment
is desired, it becomes
necessary to take into account the finite depth and the surface tension. The
main difference between surface waves in shallow water and in deeper
water lies in the fact that in the latter case dispersion effects are
important: there are two length
scales, one associated with depth and the other with surface tension,
that are responsible for wave velocity depending on wavelength. In this
paper we seek to describe the scattering of surface waves by vorticity
in terms of a single differential equation in which surface elevation is
the only dependent variable. This is possible in a perturbative
treatement away from the shallow water case, and we here present results
that correspond to first order corrections.

As in I, we consider the scattering of surface waves by a stationary
vortex, in the limit of small Mach number (the velocities of fluid
particles are small by comparison with the phase velocity of
the waves), $M \ll 1$, and large wavenumber $k$, {\it i.e.}  $\beta \equiv
k a\gg 1$ where $a$ is a typical length associated with the vortex flow.
The product $M \beta$ is assumed to be of order 1.  In Sec.
\ref{sec:interaction}, we derive, from the full hydrodynamic set of
equations, an approximation valid to order $O(M)$ (or $O(\beta^{-1})$).
First, equations are linearized for small surface perturbations around a
steady vertical vortex and then higher order terms in $M$ and $\beta^{-1}$
are discarded.  We shall pay particular attention to the orders of
magnitude of the different terms, and will justify the neglect of
dissipative effects.  The recovery of the
shallow water results is subtle since it involves taking the singular
limit of
vanishing surface tension.  There appears a partial differential equation
(Eq. \refpar{eqfin} below) that contains a squared Laplacian, and it is reduced 
to our
previous result (I-14) of ref.  [I] in the shallow water limit, i.e.  when
the layer's depth is small and surface tension is negligible.

The solution of equation \refpar{eqfin} is given in Sec.
\ref{sec:scattering}.  The results, given by \refpar{fincoeur},
\refpar{finextAB} and \refpar{finextR} and the calculations of the
Appendix, seem much more complicated than the similar shallow water
results
(I-20), (I-24) and (I-25) of [I].  However, this complexity is essentially
algebraic, and actually the physical results are rather similar, except when
dispersive effects are closely balanced by advection to yield a spiral pattern 
for the scattered waves.  The wavefront dislocation is characterized
by a parameter $\alpha$ which is a generalization of the one in [I], and
tends towards it smooothly in the shallow water limit.  In the dispersive
case, $\alpha$ depends on both the phase and group velocity of the waves.
We give a perturbative justification of the heuristic argument of Berry
{\it et al.}  \cite{berryetal}. The behaviour of the scattered wave
however depends strongly on the ratio of depth to capillary length. We also
exhibit two different behaviors,
depending on the relative values of the fluid depth and capillary length.
At each important step in the calculations, we verify that the shallow
water limit is recovered.  However, the partial differential equations
\refpar{eqfin} and (I-14) differ in the order of differentiation, with
surface tension appearing as a coefficient of the highest derivative term
in \refpar{eqfin}; the limit of null surface tension is thus singular.
Graphical illustrations of the solution are given in Sec.
\ref{sec:examples}, for various values of the dislocation parameter $\alpha$, 
and for fluid depth larger and smaller than capillary length. An 
Appendix has some computational details.

\section{Water waves in interaction with a vertical vortex}
\label{sec:interaction}

Equations for an incompressible fluid of equilibrium depth $h$,
 free surface $h + \eta (x,y,t)$ with  origin of vertical
coordinates ($z=0$) at the bottom, lying in a (uniform)
gravitational field $g$ are
\bea
\partial_t \V + \V \cdot \nabla \V & = & - \frac{1}{\rho}
\nabla P - g \hat z, \label{eq:one} \\
\nabla \cdot \V & = 0,  \label{eq:two}
\eea
where $\V$ is the fluid velocity, $P$ the pressure and $\rho$ the
(constant) density.

We neglect viscous dissipation. This is justified if the viscous
attenuation time of the wave is
greater than a typical time for the scattering problem. The attenuation
times for
gravity waves (GW) and capillary waves (CW) of wavelength $\lambda$ are,
respectively,
\cite{landau6}
\begin{equation}
T^{\rm GW} = {\rho g^2 \lambda^4 \over 2(2\pi)^4\mu \cphase^4}, \qquad
T^{\rm CW}
= {\rho  \lambda^2 \over 2(2\pi)^2\mu },
\label{attentime}
\end{equation}
where $\mu$ is the dynamic viscosity of the fluid and $\cphase$ the waves phase velocity. In the
case of water,
$\mu =
0.01\,{\rm g}/{\rm cm s}$, $g = 981\,{\rm cm}/{\rm s}^2$ and $\rho =
1\,{\rm
g}/{\rm cm}^3$. The dispersion relation for capillary-gravity waves in a
viscous fluid is fairly involved, but the sum of the two times gives a
good
estimate. Numerical estimates for waves of several wavelengths are given
in
Table
\ref{tab:OGattenuation}. The period of the wave is much smaller than the
dissipation time in all cases, and the travel time on the vortex scale,
which is
${\rm period}\times (a/\lambda)$, is also  smaller than the
attenuation time, at least as long as the vortex radius does not become
very large. Thus, there is a range of values of $ka$ where it is
reasonable to neglect dissipation.

Boundary conditions are that
fluid elements at the free surface of the  fluid remain there, that
pressure has
a discontinuity that is exactly  compensated by surface tension, and
that there
is no vertical velocity at  the bottom:
\bea
\hbox{$z=h + \eta$:} \qquad V_z & = & \partial_t \eta + {\V}_{\perp}
\cdot \nabla_\perp \eta \label{eq:bc1}\\
\hbox{$z=h + \eta$:} \qquad P & = & - \tau \nabla^2_\perp \eta
\label{eq:bc2} \\
\hbox{$z=0$:} \qquad V_z & = & 0. \label{eq:bc3}
\eea
with $\tau$ the surface tension, ${\V}_{\perp}$ the horizontal velocity
and
$\nabla_\perp$ the horizontal gradient. We are interested in small
perturbations
around a steady, axially symmetric, vertical vortex:
\bea
\label{perturb}
\V & = & \U + \pv \qquad v \ll U \\
P & = & P_0 + p_1 \qquad p_1 \ll P_0 \\
\eta & = & \eta _0 + \eta_1 \qquad \eta _1 \ll  \eta_0 .
\eea
The vertical vortex is given by the (divergenceless) flow
\beq
\label{eq:base}
\U = U_0(r) \hat{\theta}
\eeq
in cylindrical coordinates ($r$,$\theta$, $z$), with $(\hat r,
\hat{\theta}, \hat z)$ the unit vectors in the radial,
tangential and
vertical direction respectively.

We first study the zero order situation, $\pv = 0$:
\beq
\label{eq:10}
\U \cdot \nabla \U = -\frac{1}{\rho} \nabla P_0 - g\hat z .
\eeq
The $\hat{\theta}$ component of this equation is an identity. The $\hat z$
component is
\beq
0= -\frac{1}{\rho} \partial_z P_0 -g
\eeq
so that
\beq
P_0 = -\rho g z + p_0(x,y,t) ,
\eeq
and the $\hat r$ component is
\beq
\frac{U_0^2}{r} = \frac{1}{\rho} \partial_r p_0 .
\label{pressvortex}
\eeq
Given a specific function $U_0$ this is integrated at once. Concerning
boundary conditions, the third boundary condition (\ref{eq:bc3})  is
satisfied identically. The first boundary condition (\ref{eq:bc1}) says
that the surface deformation is independent of polar angle
$\theta$, and the second
boundary condition (\ref{eq:bc2}) gives the free
surface $\eta_0$ in terms of the pressure:
\beq
p_0 = \rho g \eta_0 - \tau \nabla^2_\perp \eta_0 .
\label{elevvortex}
\eeq

Writing $\pv = (\pu ,w)$ and neglecting terms quadratic in $v$ we have the
equations to order one:
\bea
(\partial_t + \U \cdot \nabla_\perp ) \pu + \pu \cdot
\nabla_\perp \U & = & - \frac{1}{\rho} \nabla_\perp p_1
\label{eqone1}\\
(\partial_t + \U \cdot \nabla_\perp ) w & = & - \frac{1}{\rho}
\partial_z p_1 \label{eqone2} \\
\nabla_\perp \cdot \pu + \partial_z w & = & 0 . \label{eqone3}
\eea
Similarly, the boundary conditions to order one are \cite{note1}
\bea
\hbox{$z=h + \eta$:} \qquad w & = & (\partial_t + \U \cdot \nabla_\perp )
\eta_1 + \pu \cdot \nabla_\perp \eta_0 \label{bcone1} \\
\hbox{$z=h + \eta$:} \qquad p_1 & = & \rho g \eta_1 - \tau \nabla^2_\perp
\eta_1 \label{bcone2} \\
\hbox{$z=0$:} \qquad w & = & 0 \label{bcone3} .
\eea
Using \refpar{eqone2} we see that the third boundary condition reads
\begin{equation}
\hbox{$z=0$:} \qquad \partial_z p_1  =  0.
\label{bcone3bis}
\end{equation}
Taking the divergence of Equations (\ref{eqone1}) and (\ref{eqone2}),
and using
(\ref{eqone3}) gives
\beq
\label{eq:div1}
\nabla_\perp^2 p_1 + \partial_{zz} p_1 = -2 \rho (\nabla_a U_b) (\nabla_b
u_a) .
\eeq
\smallskip

Up to now, these equations are exact for {\it linear} surface waves
interacting
with a static vortex. It is the fact that linear waves exist that provides
us with another parameter, the phase velocity, with which to compare $U$. 
We will now simplify the problem by using the
following two
approximations: First, the typical velocity of the vortical flow,
$U_0$, is supposed to be much less than the {\it phase} velocity of the
wave $\cphase$, defined in \refpar{defcphase}. Second, the wavelength
$\lambda$
is
supposed to be much smaller than a typical length associated with the vortex,
$a$. In
practice,
$a$ will be the core radius of the vortex, and we assume $ka\gg 1$ where
$k\equiv
2\pi/\lambda$ is the wave vector. We will denote formally the small
quantities by
$\epsilon$. We thus assume $U_0/\cphase \equiv M = O(\epsilon)$, where $M$
will
be called the Mach number in analogy with acoustics, and
$ka = O(1/\epsilon)$, and we search for corrections of order $\epsilon$ to
the
wave equation without permanent vortical flow. To get the relative
importance of
the terms that appear in the differential equations, we will use the
following estimates:
\begin{equation}
\nabla_\perp f_0 \sim {f_0\over a},\quad \partial_t f_1 \sim \nu f_1,
\quad
\nabla_\perp f_1 \sim k f_1, \quad \partial_z f_1 \sim k f_1,
\label{OG}
\end{equation}
where $f_0$ is any scalar quantity refering to the vortical flow, $f_1$
any
scalar
quantity refering to the surface waves and $\nu$ is the wave frequency.
We have assumed, as suggested by Eqn. (\ref{eq:div1})
that length scales for vertical
and horizontal variations of surface waves are the same.

With those estimates, we get from \refpar{eqone2} that
\begin{equation}
{k\over \rho}p_1 \sim \nu w \qquad\Longrightarrow\qquad p_1 \sim \rho
\cphase w.
\label{OGp}
\end{equation}
Injecting this result in \refpar{eq:div1}, the order of magnitude of the
left-hand side is $k^2p_1 = k^2\rho \cphase w$, whereas the
right-hand
side is of order $\rho (U_0/a)kw = k^2\rho \cphase w (U_0/\cphase)(1/ka)$; 
it is thus negligible, being of order $O(\epsilon^2)$,
and
\refpar{eq:div1} is replaced by
\begin{equation}
\nabla_\perp^2 p_1 + \partial_{zz} p_1 = 0,
\label{laplace}
\end{equation}
which is the same Laplace equation as in the problem of water waves
without
the
vortex; it has the big advantage of being autonomous and linear in the
pressure so that separation of variables can be attempted.

 An estimate of the surface elevation $\eta_0$ for the vortex flow,
may be
obtained from \refpar{elevvortex} and \refpar{pressvortex}; it reads
\begin{equation}
\eta_0 \sim {U_0^2 \over g}\left(1 + {l_c^2\over a^2}\right)^{-1},
\label{OGetaV}
\end{equation}
where we have introduced the {\it capillary length} $l_c \equiv
\sqrt{\tau/\rho
g}$. For water, $\tau = 74\,{\rm dyn}/{\rm cm}$, so that $l_c \approx
0.32{\rm cm}$ and
the effect of surface tension on the surface deformation of a
vortex of size $a \approx 1$ cm, is quite small, of
order one per cent. The surface wave
elevation, from \refpar{bcone2} and \refpar{OGp}, reads
\begin{equation}
\eta_1 \sim {\cphase w \over g}\left(1 + k^2 l_c^2\right)^{-1},
\label{OGetaW}
\end{equation}
with $k^2 l_c^2$ of order one. In the following, we take $\eta_1 \sim
{\cphase w
/ g}$, which is numerically inexact but adequate for order of
magnitude considerations. In \refpar{bcone1}, the respective orders of
magnitude of
the different terms are
\begin{eqnarray}
\partial_t \eta_1 &\sim & {k \cphase^2 \over g}w , \nonumber \\
\quad \U \cdot
\nabla_\perp \eta_1 &\sim & {k \cphase^2 \over g}\;{U_0 \over \cphase} w =
O(\epsilon ) (\partial_t \eta_1), \nonumber \\
\pu \cdot \nabla_\perp \eta_0 &\sim & {k \cphase^2 \over g} \; {1\over ka}
{U_0^2\over \cphase^2}w = O(\epsilon^3 ) (\partial_t \eta_1),
\label{OGprep}
\end{eqnarray}
and the relevant approximation for \refpar{bcone1}, valid to
$O(\epsilon)$,
reads
\begin{equation}
\hbox{$z=h$:} \qquad w  =  (\partial_t + \U \cdot \nabla_\perp )
\eta_1 .
\label{approxbcone1}
\end{equation}
In the same approximation, we also write
\begin{equation}
\hbox{$z=h$:} \qquad p_1  =  \rho g \eta_1 - \tau \nabla^2_\perp
\eta_1
\label{approxbcone2}
\end{equation}

In those equations, we neglect $\eta$ in comparison with $h$.
If we write $p_1(h
+ \eta_0) = p_1(h) + \delta p_1$ and use \refpar{OGetaV} we get
\begin{equation}
{\delta p_1 \over p_1}  = {\eta_0 \over p_1}\left.{{\rm d} p_1 \over {\rm
d}
  z}\right|_h \sim k\eta_0 \sim k{U_0^2 \over g} \sim {U_0^2 \over
\cphase^2} =
O(\epsilon^2),
\label{OGbound}
\end{equation}
so that $\delta p_1$ is indeed negligible. The same is true for the other
boundary equation.

Let us use now these approximate equations to describe the propagation of
surface
waves in the vortical flow. We will consider almost shallow water waves,
that
is, the next order in the small parameter $kh$ of the calculations of ref.
(I). In
this limit, the pressure is given as a power series in the vertical
coordinate
$z$,
\begin{equation}
p_1(r,\theta,z,t) = \sum_{n = 0}^{\infty}z^n \Pi_n(r,\theta,t).
\label{pressionform}
\end{equation}
A proper development would consider the dimensionless variable $z/L$ with
$L$ a typical length scale for horizontal variations. However, 
(\ref{pressionform}) is good enough for our purposes. Should we wish to
have exact results in the deep water limit, $kh \rightarrow \infty$, this
is the step that would break down.
Inserting this development in \refpar{laplace}, we obtain the recursion
\begin{equation}
\Pi_{n+2} = {-\nabla_\perp^2 \Pi_n \over (n+2)(n+1)},
\label{recursion}
\end{equation}
which, together with the boundary condition \refpar{bcone3bis}, gives
\begin{equation}
p_1(r,\theta,z,t) = \sum_{m = 0}^{\infty}(-1)^m z^{2m} {\nabla_\perp^{2 m}
\Pi
\over (2 m)!}.
\label{pression}
\end{equation}
where we have dropped the index `0' from $\Pi_0$ for convenience.

We introduce the notation $D_t \equiv \partial_t +
\U \cdot \nabla_\perp$. Taking only the leading order terms in the small
parameter $kh$ in
\refpar{pression}, applying the differential operator $D_t$ to
\refpar{approxbcone1} and taking Eqn.
\refpar{eqone2} for
$z = h$, we get
\begin{equation}
D_t^2 \eta_1 = {h\over \rho}\nabla_\perp^2\Pi - {h^3\over
6\rho}\nabla_\perp^4\Pi.
\label{eqinter1}
\end{equation}
Applying $\nabla_\perp^2$ to \refpar{approxbcone2} and using
\refpar{pression} at the same order, we get
\begin{equation}
  g h\nabla_\perp^2 \eta_1 - {\tau h \over \rho} \nabla_\perp^4 \eta_1 =
{h\over
\rho}\nabla_\perp^2\Pi - {h^3\over 2\rho}\nabla_\perp^4\Pi.
\label{eqinter2}
\end{equation}
The surface tension term is considered under the assumption that the
capillary
length is of the same order of magnitude as the depth of the fluid layer.
In
this case, those two equations are valid up to order
$O[(kh)^2]$. It is thus legitimate to replace the pressure by its value at
order
$O(1)$,
$\Pi = \rho g \eta_1$, in the term $\propto \nabla_\perp^4\Pi$, that has
the highest derivative.
Eliminating the pressure in the resulting equations, we get the final
result: a dispersive wave equation for surface elevation $\eta_1$
that is analogous to Eqn. (I-14) of I in the shallow water case. It reads
\begin{equation}
gh \nabla_\perp^2 \eta_1 + \left({1\over 3}gh^3 - { \tau h \over
\rho}\right)\nabla_\perp^4
\eta_1  - D_t^2 \eta_1 = 0.
\label{eqfin}
\end{equation}
This equation includes the leading order correction to the shallow water
case. It is
valid
under the same assumptions (see ref.
[I]) concerning wavelength and fluid velocity. It describes the scattering
of surface waves over water whose depth is small but not negligible
with
respect to wavelength, when the wavelength is small
compared to the vortex size, when the velocity of the vortex flow is much
less
than the phase velocity of the waves, and when the waves are of small
amplitude.

Without the vortex, when $U = 0$ and $\partial_t = D_t$,
plane progressive waves of the form
$$\eta_1 \propto e^{i(\nu t - {\boldmath k}_\perp \cdot{\boldmath
r}_\perp)}$$
exist provided frequency $\nu$ and wave vector ${\boldmath k}_{\perp}$
are related through the dispersion relation
\begin{equation}
\nu^2 = ghk^2 + \left( { \tau h \over
\rho} - {1\over 3}gh^3 \right)k^4 .
\label{reldispapprox}
\end{equation}
The phase velocity $\cphase$ is given by
\begin{equation}
\cphase^2 = {\nu^2 \over k^2} = gh + \left( { \tau h \over
\rho} - {1\over 3}gh^3 \right)k^2,
\label{defcphase}
\end{equation}
and the group velocity $c_{\rm g}$ by
\begin{equation}
c_{\rm g} = {1\over \cphase}\left[gh + 2\left( { \tau h \over
\rho} - {1\over 3}gh^3 \right)k^2\right].
\label{defcgroupe}
\end{equation}
Of course, the dispersion relation \refpar{reldispapprox} is the
approximation to
order $O[(kh)^3]$ of the well known dispersion relation for
capillary-gravity waves,
\begin{equation}
\nu^2 = \left(gk + {\tau k^3 \over \rho}\right)\tanh kh.
\label{reldisp}
\end{equation}

The wave dispersion is thus characterized by a dimensionless parameter
$\delta$ defined by
\begin{equation}
k^2\left({\tau \over \rho g} - {h^2 \over 3}\right) = {1\over \delta}.
\label{defdelta}
\end{equation}
It is positive for  $h < \sqrt{3} l_c$, and negative otherwise. We shall
consider both cases. In order to be consistent with
our approximations, namely, that the fourth order term in (\ref{eqfin}) be
a small correction to the other two, the
absolute value of $\delta$ must be large, and the shallow
water limit corresponds to $|\delta | \to \infty$. For positive
$\delta$, the phase and group velocity  read
\begin{equation}
\cphase^2 = gh \, {1 + \delta \over \delta}, \qquad c_{\rm g} = {gh \over
\cphase}{2 + \delta \over \delta},\qquad (\delta > 0),
\label{vitdeltapos}
\end{equation}
whereas for negative values of $\delta$ they read
\begin{equation}
\cphase^2 = gh \, {|\delta | -1\over |\delta |}, \qquad c_{\rm g} = {gh
\over
\cphase}{| \delta | -2 \over |\delta |},\qquad (\delta < 0).
\label{vitdeltaneg}
\end{equation}

The full dispersion relation \refpar{reldisp} for water waves is either
convex or
concave at small depth, depending on the sign of
$\delta$. The crossover point $h = \sqrt{3} l_c$, derived from the approximate
relation \refpar{reldispapprox}, separates two regions of opposite convexity.
Both may be experimentally accessible. The approximation of the hyperbolic
tangent is better than $1\%$ for $kh < 0.5$, and better than $5\% $ for $kh <
0.8$. It is thus easy to stay in the small depth limit, $\tanh (kh)
\approx kh - (kh)^3/3$, while keeping the wavelength small in comparison
with the
vortex radius. Using a fluid with high surface tension, like water, leads to a
positive
$\delta$, that is $h <\sqrt{3} l_c$, whereas the same experiment with a
fluid of
small surface tension will give a negative $\delta$.

\section{Scattering of dislocated waves by a vortex}
\label{sec:scattering}

We will now proceed in close analogy with the calculations of ref. [I]. We
consider
scattering of surface waves by a circular uniform vortex with vorticity
$\omega$ and radius $a$ surrounded by an irrotational flow.
Using polar coordinates $(r, \theta)$, the background flow is given
by\cite{foot1new}
\begin{equation}
\mbox{\boldmath $U$} =
\left\{
\begin{array}{lcl}
\frac{1}{2} \omega r \hat{\theta}  &  & \mbox{if  $r\le a$} \\
 & & \\
\frac{\Gamma}{2\pi r} \hat{\theta} &  & \mbox{if  $r > a$}
\end{array}
\right.
\label{v1}
\end{equation}
where $\Gamma = \pi \omega a^2$ is the circulation.
Eqn. (\ref{eqfin}) will be solved separately for $r<a$ and $r>a$, and the
results matched with a continuity condition for $\eta_1$ and its first
three
derivatives, since \refpar{eqfin} is of order four.

We look for solutions that evolve harmonically (with a single global
frequency $\nu$) in time, and Fourier decompose them in the polar angle
$\theta$:
\begin{equation}
\displaystyle
\eta_1 = {\rm Re} \left[ \sum_n \widetilde{\eta}_{1n} {\rm e}^{i (n \theta
- \nu
 t )} \right] ,
\label{decomp}
\end{equation}
where Re stands for the real part. Inserting this development, and the
background
flow \refpar{v1} for $r \le a$, that is, inside the vortex,
in \refpar{eqfin},
a little calculation shows that the resulting equation factorizes exactly
as
\begin{equation}
\left[{{\rm d}^2 \over {\rm d} r^2} + {1\over
r}{{\rm d}\over {\rm d} r} - {n^2\over r^2}  +
(k_+)^2\right]\left[{{\rm d}^2 \over {\rm d} r^2} + {1\over
r}{{\rm d}\over {\rm d} r} - {n^2\over r^2}  +
(k_-)^2\right]\widetilde{\eta}_{1n} = 0,
\label{scattcoeur}
\end{equation}
with
\begin{equation}
(k_\pm)^2 \equiv {1 \over 2}k^2 \delta\left(-1 \pm \sqrt{1 +
{4\left(\nu - {n\omega / 2}\right)^2 \over ghk^2\delta}}\right),\qquad
(\delta >
0),
\label{kplusmoinspos}
\end{equation}
or
\begin{equation}
(k_\pm)^2 \equiv {1 \over 2}k^2 |\delta |\left(1 \pm \sqrt{1 -
{4\left(\nu - {n\omega / 2}\right)^2 \over ghk^2|\delta |}}\right),\qquad
(\delta
< 0),
\label{kplusmoinsneg}
\end{equation}

The two differential operators in \refpar{scattcoeur} commute, and the
four
independent solutions of this fourth order equation
are thus given by the two pairs of solutions of  the two corresponding
second order differential equations.

Taking the shallow water limit $\delta \to \infty$, we get for positive
$\delta$
\begin{equation}
(k_+)^2 \to {\left(\nu - {n\omega \over 2}\right)^2\over gh},
\label{kplusmoinsapprox}
\end{equation}
whereas $k_-$ diverges. Thus $k_+$ tends towards the value of $k_n$
corresponding to the
shallow water case (see Eqn. (19) of ref. [I]) as it should, since this
case
must be recovered as a limiting case. The other
constant $k_-$ comes from the fact that \refpar{scattcoeur} is a fourth
order
differential equation, unlike Eqn. (19) of ref. [I]. Its limit for $\delta
\to
\infty$ is singular, reflecting the fact that surface tension $\tau$
multiplies
the highest derivative term in the
differential equation (\ref{eqfin}). The respective role of $k_+$ and
$k_-$
are
exchanged for negative $\delta$.

From \refpar{kplusmoinspos}, we see that when $\delta$ is positive $k_+$
is
real
whereas
$k_-$ is imaginary for all $n$, whereas for negative $\delta$ the two
wavevectors $k_\pm$ are real for small $n$ and complex for large $n$. For
positive
$\delta$,
\refpar{scattcoeur} has Bessel and Neumann functions as solutions,
together
with
hyperbolic Bessel and hyperbolic  Neumann functions. The Neumann and
hyperbolic Neumann functions must be discarded because of
regularity at the origin. For negative $\delta$, we take Bessel and
Neumann
functions of a complex argument, and discard the Neumann functions to
ensure
regularity at the origin. Thus
\begin{equation}
\eta_1 = {\rm Re} \left[\sum_n \left(a_n {J_n(k_n r) \over J_n(k_n a)} +
 b_n {X_n(\kappa_n r) \over X_n(\kappa_n a)}\right){\rm e}^{i (n
\theta -
\nu
 t )} \right] ,
\label{fincoeur}
\end{equation}
where the $a_n$ and $b_n$ are undetermined coefficients in both cases, and
where
we  have introduced
the notation
\begin{equation}
k_+ \equiv k_n, \quad k_- \equiv i\kappa_n, \quad X_n \equiv I_n, \qquad
(\delta
> 0),
\label{defcoeurpos}
\end{equation}
\begin{equation}
k_- \equiv k_n, \quad k_+ \equiv \kappa_n, \quad X_n \equiv J_n, \qquad
(\delta
< 0).
\label{defcoeurneg}
\end{equation}

Outside the vortex, for $r>a$, using $\mbox{\boldmath $U$} =\Gamma/(2\pi
r)
\hat{\theta}$ and the decomposition \refpar{decomp}, and dropping terms of
order
$M^2$, we get that
\refpar{eqfin} reads
\begin{eqnarray}
\left[\left({\cal L} - {n^2 \over r^2}\right)^2 - \delta k^2\left({\cal L}
- {n^2
\over r^2}\right)  - A + {B\over r^2}
\right]\widetilde{\eta}_{1n} = 0,
\label{extdebut}
\end{eqnarray}
where, using the dispersion relation \refpar{reldispapprox}, we define the
two
constants
\begin{equation}
 A \equiv (\delta +1)k^4, \quad B
\equiv {\delta k^2 \over gh}\,{2\Gamma \nu n \over 2 \pi},
\label{defconstant}
\end{equation}
and the linear differential operator
\begin{equation}
{\cal L} \equiv {{\rm d}^2 \over {\rm d} r^2} + {1\over
r}{{\rm d}\over {\rm d} r}.
\end{equation}

This equation may be written in the factorized form
\begin{equation}
O_+ O_- \widetilde{\eta}_{1n} = 0,
\label{extinter}
\end{equation}
with
\begin{equation}
O_\pm \equiv {\cal L} - {m_{\pm}^2\over r^2}  +
q_{\pm}^2 ,
\end{equation}
provided the unknown coefficients $m_+$, $m_-$, $q_+$ and $q_-$
satisfy the following relations:
\begin{eqnarray}
({\cal L}): \quad & \Longrightarrow & \quad q_+^2 + q_-^2 = -\delta k^2,
\label{rel1}
\\ (1): \quad & \Longrightarrow & \quad q_+^2  q_-^2 = -A, \label{rel2} \\
\left({1 \over r^2}\right): \quad & \Longrightarrow & \quad m_+^2q_-^2 +
m_-^2q_+^2 = -\delta k^2 n^2 - B, \label{rel3} \\
\left({{\cal L} \over r^2}\right): \quad & \Longrightarrow & \quad m_+^2 +
m_-^2 = 2n^2, \label{rel4} \\
\left({1 \over r^3}{{\rm d} \over {\rm d}r}\right): \quad &
\Longrightarrow
&
\quad  m_-^2 = n^2, \label{rel5} \\
\left({1 \over r^4}\right): \quad & \Longrightarrow & \quad m_+^2  m_-^2
-
4
m_-^2 = n^4 - 4 n^2 . \label{rel6}
\end{eqnarray}
Here we have indicated on the left the portion of the linear differential
operator that leads to each condition. There are six equations for only four 
unknowns and obviously
they cannot be simultaneously satisfied in general. The last two
equations, (\ref{rel5}) and (\ref{rel6}), correspond to
terms that are negligible at large distance from the vortex. If we compare
them
to ${\cal L}^2$, they are smaller than $1/\beta^2$ because $r>a$.
Accordingly, we are justified in ignoring these two equations in our
approximation $\beta \gg 1$ and we solve
(\ref{rel1}-\ref{rel4}), which gives
\begin{equation}
q_+^2 = k^2 > 0, \quad q_-^2 \equiv (iq)^2 = - k^2(1+\delta) < 0,\quad
(m_\pm)^2
= n^2 \pm 2 n \alpha,\qquad (\delta > 0),
\label{extqplusmoinspos}
\end{equation}
\begin{equation}
q_+^2 \equiv q^2 = (|\delta |-1)k^2 > 0, \quad q_-^2 =  k^2 > 0,\quad
(m_\pm)^2
= n^2 \mp 2 n \alpha,\qquad (\delta < 0),
\label{extqplusmoinsneg}
\end{equation}
\begin{equation}
  \alpha \equiv {\Gamma \nu \over 2 \pi}\,{1
\over gh + 2(\tau/\rho - gh^2/3)hk^2} = {\Gamma \nu \over 2 \pi}\,{1
\over \cphase c_{\rm g}},
\label{defalpha}
\end{equation}
where we used \refpar{defcgroupe} to write the last equality. It is
important to
note that the index $n^2 + 2n\alpha$ is always associated to the
incident wavevector
$k$. We will comment further on this result after Eqn. \refpar{finextAB}. From 
now on, we define
\begin{equation}
m_+ \equiv \sqrt{n^2 + 2 n \alpha}, \qquad m_- \equiv \sqrt{n^2 - 2 n
\alpha},
\label{defmplusfinale}
\end{equation}
so that in the negative $\delta$ case $m_-$ (resp. $m_+$) is associated
with
$q_+$ (resp. $q_-$), as shown by \refpar{extqplusmoinsneg}.

The dimensionless parameter $\alpha$ is defined in close analogy with I.
We can write $\alpha = M\beta (\cphase/c_{\rm g})$, which may be of order
1,
whith
$M\ll 1$ and $\beta \gg 1$. This parameter has the same physical
interpretation
as in the shallow water case (see below) : it gives the amount of
dislocation for
the wavefronts far from the vortex. This calculation provides an
explicit confirmation, in a
perturbation expansion near the shallow water case, of the intuitive result of
Berry
{\it et al.} \cite{berryetal}.

The two differential operators $O_\pm$ in \refpar{extinter} do not
commute.
Using
the usual notation $[\cdot ,\cdot ]$ for the commutator of two operators,
we get
\begin{equation}
[O_+,O_-] = \left(m_+^2 - m_-^2\right)\left[{\cal L},{1\over r^2}\right] =
4\left(m_+^2 - m_-^2\right)\left({1\over r^4} - {1\over r^3}{{\rm d} \over
{\rm
d}r}\right),
\label{commutator}
\end{equation}
which is small, of the same order as the neglected terms (\ref{rel5},
\ref{rel6}),
and will also be neglected. Thus, in the
same approximation, for positive $\delta$ the solution of
\refpar{extinter}
is a
linear combination of Bessel, Neumann, hyperbolic Bessel and hyperbolic
Neumann
functions, because
$q_+$ is real and $q_-$ is
imaginary. Since the hyperbolic Bessel function diverges at
infinity, it must be discarded. For negative $\delta$, the solution is a
linear
combination of Bessel and Neumann functions of argument $kr$ and $qr$.
Since the
wave number $q$ is that of a scattered wave, we discard the Bessel
function
of
$qr$, keeping only the outgoing wave from the vortex.

Following Berry {\it et al.}\cite{berryetal} as
in [I], we write the surface elevation outside the vortex in the form
\begin{equation}
\eta_1 = {\rm Re} (\eta_{AB}+\eta_R),
\label{v8}
\end{equation}
where
\begin{equation}
\eta_{AB}= \sum_n c_n {J_{m_+}(k r) \over J_{m_+}(\beta)}
{\rm e}^{i (n
\theta -
\nu
 t )}  ,
\label{finextAB}
\end{equation}
 with $\beta \equiv ka$. It does not depend on the sign of $\delta$, which
is
physically obvious because the amount of dislocated wavefront is linked to
the
circulation of the vortex, not to the curvature of the dispersion
relation. Thus $m_+ = \sqrt{n^2 + 2n\alpha}$ is always the index of the
functions involving the wavevector $k$. The other term of \refpar{v8}
depends on
the sign of $\delta$, which is also physically clear since they represent
the
wave scattered by the vortex. They read
\begin{equation}
\eta_{R}= \sum_n \left( d_n {H_{m_+}^1(k r) \over H_{m_+}^1(k a)} +
e_n{Y_{m_-}(q r) \over Y_{m_-}(q a)} \right){\rm e}^{i (n
\theta -
\nu
 t )}  ,
\label{finextR}
\end{equation}
 Depending on the sign
of $\delta$, we have the following definitions :
\begin{equation}
Y_{m_-} \equiv K_{m_-}, \quad q = k\sqrt{1 + \delta }, \qquad (\delta > 0)
\label{defextpos}
\end{equation}
\begin{equation}
Y_{m_-} \equiv H^{1}_{m_-}, \quad q = k\sqrt{|\delta | - 1}, \qquad
(\delta
< 0)
\label{defextneg}
\end{equation}

The coefficients
$a_n$,
$b_n$, $c_n$, $d_n$ and $e_n$ are defined so that they denote the
amplitude
of
the wave components at the vortex boundary $r=a$. In order to obtain these
coefficients, and since the equation \refpar{eqfin} is of order four, the
continuity of
$\widetilde{\eta}_1$ and its first three derivatives at $r=a$ is required.
This gives
four relations:
\begin{equation}
a_n + b_n - d_n - e_n = c_n ,
\label{rela}
\end{equation}
\begin{equation}
a_n k_n {J_n'(k_n a) \over J_n(k_n a)} +
 b_n \kappa_n{X_n'(\kappa_n a) \over X_n(\kappa_n a)} - d_n
k{{H_{m_+}^1}'(\beta)
\over H_{m_+}^1(\beta)} - e_n q{Y_{m_-}'(q a) \over Y_{m_-}(q a)} = c_n k
{J_{m_+}'(\beta) \over J_{m_+}(\beta)},
\label{relb}
\end{equation}
\begin{equation}
a_n k_n^2 {J_n''(k_n a) \over J_n(k_n a)} +
 b_n \kappa_n^2{X_n''(\kappa_n a) \over X_n(\kappa_n a)} - d_n
k^2{{H_{m_+}^1}''(\beta)
\over H_{m_+}^1(\beta)} - e_n q^2{Y_{m_-}''(q a) \over Y_{m_-}(q a)} = c_n
k^2
{J_{m_+}''(\beta) \over J_{m_+}(\beta)},
\label{relc}
\end{equation}
\begin{equation}
a_n k_n^3 {J_n'''(k_n a) \over J_n(k_n a)} +
 b_n \kappa_n^3{X_n'''(\kappa_n a) \over X_n(\kappa_n a)} - d_n
k^3{{H_{m_+}^1}'''(\beta)
\over H_{m_+}^1(\beta)} - e_n q^3{Y_{m_-}'''(q a) \over Y_{m_-}(q a)} =
c_n
k^3
{J_{m_+}'''(\beta) \over J_{m_+}(\beta)},
\label{reld}
\end{equation}
where $X$ (resp. $Y$) is defined either by \refpar{defcoeurpos} or
\refpar{defcoeurneg} [resp. \refpar{defextpos} or \refpar{defextneg}].

The fifth, and last, boundary condition comes from the asymptotic
behaviour of $\eta$
at
infinity.  We require that the asymptotics of $\eta_{AB}$ coincides with
the dislocated wave incident from the right plus outgoing waves only.
Exactly in
the same way as in I, this leads to
\begin{equation}
\frac{c_n}{J_{m_+}(\beta)} = (-i)^{m_+},
\label{v15}
\end{equation}
It is important, in order to use this result, that either the
coefficient $q_+^2$, for positive $\delta$, or $q_-^2$, for negative
$\delta$, in
\refpar{extinter} be equal to
$k^2$, and that they be associated in each case to $m_+$. Otherwise it would
have
been impossible to recover the dislocated wave, which is a crucial
physical
requirement for the solution because we need $q=k$ to be a possible result
of the factorization. This fact fully justifies the factorization
in
\refpar{extinter}.

The solution of system (\ref{rela}--\ref{reld}) is thus known in
principle,
but it is not too illuminating, and it will not be displayed here.
Details of the
calculations may be found in the Appendix. To illustrate the
solution, we use Mathematica
\cite{mathematica} in order to do the
calculations that will be indicated in Sec. \ref{sec:examples}. Thus it is 
sufficient to have the
coefficients expressed as ratio of $4\times 4$ determinants, as in
\refpar{anformel}. We insist on the fact that the solution may be
innacurate at a
few wavelengths  away from the vortex, because of the approximate character of
the
factorization \refpar{extinter}.

Let us discuss the asymptotic behavior
of the solution (\ref{finextAB}, \ref{finextR}) for $r\to \infty$. The
case
of
$\eta_{AB}$ is completely similar to the shallow water case. Indeed, the
index of
the Bessel function, $m_+(n) = \sqrt{n^2 + {\rm (Const.)}\times n}$, has
exactly the
same structure as $m(n)$ in [I]. An important consequence is that the
parameter
$\alpha = \beta M(\cphase/c_{\rm g})$ has the same physical significance
as in the shallow water (or acoustics) case : it quantifies the dislocation of
the wavefronts in the forward direction, at large distances from the vortex.
Other
results may also be transposed in a straightforward fashion, and the asymptotics
of $\eta_{AB}(r,\theta)$ for large $r$ is still given by Eqn. (38) of [I],
with the proviso that
the function $G(\theta, -\pi/2)$ be replaced by
\begin{equation}
G_{DW}(\theta, t) \equiv \sum_{|n|<N}{\rm e}^{in\theta}\left[{\rm
e}^{im_+(t -
\pi/2)} - {\rm e}^{im_{\rm old}(t -
\pi/2)}\right],
\label{defGDW}
\end{equation}
where $m_+$ is given by \refpar{defmplusfinale} and
\begin{equation}
m_{\rm old} = \left|n + {\Gamma \nu \over 2 \pi}\,{1
\over \cphase c_{\rm g}}\right|.
\label{expmold}
\end{equation}

The asymptotics of $\eta_R$ depends on the sign of $\delta$. If $\delta$
is
positive, the hyperbolic Bessel function does not contribute to the
scattered far
field, because
\cite{grad} $K_p(z) \sim {\rm e}^{-z}/\sqrt{z}$ for large $z$. Thus
\begin{equation}
\label{v17}
\eta_R  \rightarrow \left(\frac{2}{\pi i k r}\right)^{1/2}
{\rm e}^{i(kr-\nu t)}
\sum_n {d_n \over H_{m_+}^1(\beta)}{\rm e}^{i(n\theta -\pi m_+/2)}.
\end{equation}
In the next section, we will compare the
correction to the Aharonov-Bohm scattering amplitude in the case of
shallow
water waves, given by Eqn. (43) of [I], and the correction for dispersive
water
waves, which reads
\begin{equation}
\widetilde{f}_{DW}(\theta )  =   G_{DW}(\theta , -\pi /2)  +
2 \sum_n \frac{d_n}{H_{m_+}^1(\beta)}  {\rm e}^{in \theta }(-i)^{m_+}.
             \label{v20pos}
\end{equation}

If $\delta$ is negative, we must take into account the two outgoing Hankel
functions. We get
\begin{equation}
\label{v17neg}
\eta_R  \rightarrow \left(\frac{2}{\pi i k r}\right)^{1/2}
{\rm e}^{i(kr-\nu t)}
\sum_n \left[{d_n {\rm e}^{-i\pi m_+/2}\over H_{m_+}^1(\beta)}
+ {e_n {\rm
e}^{-i\pi m_-/2}\over(|\delta| -1)^{1/4}
H_{m_-}^1\left(\beta\sqrt{|\delta|
-1}\right)}\right]{\rm e}^{in\theta},
\end{equation}
and the correction for dispersive water
waves now reads
\begin{equation}
\widetilde{f}_{DW}(\theta )  =   G_{DW}(\theta , -\pi /2)  +
2 \sum_n \left[\frac{d_n (-i)^{m_+}}{H_{m_+}^1(\beta)} + \frac{e_n
(-i)^{m_-}}{ (|\delta| -1)^{1/4}H_{m_-}^1\left(\beta\sqrt{|\delta|
-1}\right)} \right]{\rm e}^{in
\theta },
             \label{v20neg}
\end{equation}

\section{Numerical Examples}
\label{sec:examples}

The solution to the scattering problem of surface waves by a uniform
vertical vortex depends on four
dimensionless parameters. A first set includes the dimensionless vortex
radius
$\beta \gg 1$ and the dislocation parameter $\alpha = \beta
M(\cphase/c_{\rm g})$ which
quantifies the wavefront dislocation. They already appeared in [I], with
the same
definitions and physical interpretations. A third parameter is the
dimensionless capillary length
$\ell \equiv k l_c$, and the last one is the dimensionless depth
$kh$. In order to simplify somewhat the discussion, we use the single
dimensionless parameter $\delta$, defined in \refpar{defdelta}, in  place
of
the depth and capillary length. As an example, we take
$\delta = 8$, which may correspond, for example, to
$h = l_c$, $kh = \sqrt{3}/4$, and  $\delta = -8$, which may correspond to
$h = 3 l_c$, $kh = 3/4$. In both cases, the
hyperbolic tangent in  (\ref{reldisp}) is
approximated to better than five percent by the two leading terms, the
ones
we are keeping, in its series expansion.

Scaling radial distance with the vortex radius, $r' \equiv r/a$,
the analytical expression of the surface displacement is summarized as
follows, depending on the sign of $\delta$. Inside the vortex ($0<r'\le
1$) we have
$$
\eta_1 = {\rm Re} \ \eta_c.
$$
For {\it positive} values of $\delta$,
\begin{equation}
\eta_c  =  \sum_{n } \left(a_n \frac{J_{n}(\tilde{\phi}_n 
r')}{J_{n}(\tilde{\phi}_n )} +
b_n \frac{I_{n}(\varphi_n r')}{I_{n}(\varphi_n )}\right)
 {\rm e}^{i(n\theta -\nu t)}  ,
\label{v27pos}
\end{equation}
where we have defined the following dimensionless wave numbers:
\begin{equation}
k_n a \equiv \tilde{\phi}_n = \beta \sqrt{{\delta\over 2}}\left(-1 + \sqrt{1
+ 4{1 + \delta\over \delta^2}\left(1 - n{\alpha\over \beta^2}{2 +
\delta\over
1 + \delta}\right)^2}\right)^{1/2},
\label{adimk1pos}
\end{equation}
\begin{equation}
\kappa_n a \equiv \varphi_n = \beta \sqrt{{\delta\over 2}}
\left(1 +
\sqrt{1 + 4{1 + \delta\over \delta^2}\left(1 - n{\alpha\over \beta^2}{2 +
\delta\over 1 + \delta}\right)^2}\right)^{1/2}.
\label{adimk2pos}
\end{equation}
For {\it negative} values of $\delta$,
\begin{equation}
\eta_c  =  \sum_{n } \left(a_n \frac{J_{n}(\phi_n r')}{J_{n}(\phi_n )} +
b_n \frac{J_{n}(\tilde{\varphi}_n r')}{J_{n}(\tilde{\varphi}_n )}\right)
 {\rm e}^{i(n\theta -\nu t)}  .
\label{v27neg}
\end{equation}
whith new dimensionless wave numbers:
\begin{equation}
k_n a \equiv \phi_n = \sqrt{{|\delta|\over 2}}\beta \left(1 -
\sqrt{1 - 4{ |\delta| - 1\over \delta^2}\left(1 - n{\alpha\over \beta^2}{
|\delta| - 2\over  |\delta| - 1}\right)^2}\right)^{1/2},
\label{adimk1neg}
\end{equation}
\begin{equation}
\kappa_n a \equiv \tilde{\varphi}_n = \sqrt{{|\delta|\over 2}}\beta
\left(1 +
\sqrt{1 + 4{ |\delta| - 1\over \delta^2}\left(1 - n{\alpha\over \beta^2}{
|\delta| - 2\over |\delta| - 1}\right)^2}\right)^{1/2}.
\label{adimk2neg}
\end{equation}

Outside the vortex ($r'>1$)
$$
\eta_1 = {\rm Re} (\eta_{AB} + \eta_{R}),
$$
where, whatever the sign of $\delta$,
\begin{equation}
\eta_{AB} = \sum_n (-i)^{m_+} J_{m_+}(\beta r') {\rm e}^
{i(n\theta -\nu t)}.
\label{v28}
\end{equation}
For {\it positive} values of $\delta$,
\begin{equation}
\eta_R = \sum_n \left(d_n \frac{H^1_{m_+} (\beta r')}{H^1_{m_+} (\beta)} +
e_n \frac{K_{m_-} \left(\beta \sqrt{1 + \delta}
r'\right)}{K_{m_-}\left(\beta
\sqrt{1 + \delta}\right)}
\right){\rm e}^{i(n\theta -\nu t)}.
\label{v29pos}
\end{equation}
where we have used the fact that
\begin{equation}
q a  = \beta \sqrt{1 + \delta}.
\label{adimk3pos}
\end{equation}
For {\it negative} values of $\delta$,
\begin{equation}
\eta_R = \sum_n \left(d_n \frac{H^1_{m_+} (\beta r')}{H^1_{m_+} (\beta)} +
e_n \frac{H^1_{m_-} \left(\beta \sqrt{|\delta| - 1}
r'\right)}{H^1_{m_-}\left(\beta
\sqrt{|\delta| - 1}\right)}
\right){\rm e}^{i(n\theta -\nu t)}.
\label{v29neg}
\end{equation}
where we have used the fact that
\begin{equation}
q a  = \beta \sqrt{|\delta| - 1}.
\label{adimk3neg}
\end{equation}
Some details of the calculations of the coefficients $a_n$, $b_n$, $d_n$
and
$c_n$ are given in the Appendix. As an illustration, absolute values of
the
coefficients $a_n$ and $b_n$ (resp. $d_n$ and $e_n$) are plotted in Fig. 1
(resp. in Fig. 2), for positive $\delta$, and in Figs. 3 and 4 for
negative $\delta$.

Since convergence of the series expansions for $\eta_{AB}$ and $\eta_R$ is
not uniform, the number of terms to keep in the numerical evaluation of the 
infinite series depends on the
value of $r'$. In practice, the convergence of the series is comparable to
the
case of ref. [I], and we use roughly the same number of terms. We compute
the
patterns of the surface displacement in the
region $|x'|, |y'| \le 5  [ (x',y')=(r'\cos \theta, r' \sin \theta )]$ by
the
finite  sum of (\ref{v27pos}, \ref{v27neg}) and (\ref{v29pos},
\ref{v29neg})
with $|n| \le 50$ for $\beta =10$ and
$|n|
\le 30 $ for $\beta =5$, but we keep more terms, $|n|
\le 90 $ in \refpar{v28}.

Let us first consider the case of positive
$\delta$. Fig. 5 shows the
resulting
displacements for $\delta = +8$, $\beta = 5$ and
$\alpha = 0.5,\,1,\,1.5,\,2$, and Fig. 6 for the same values of $\alpha$ and 
$\delta$, but 
$\beta = 10$. The dislocation of the
incident
wavefronts by an amount equal to $\alpha$ is clearly visible. The outward
travelling  scattered wave is also visible. The interference
patterns between scattered and incident wave is very similar to the
corresponding pictures of [I], for the same values of $\beta$ and
$\alpha$.
This
is confirmed by the comparison of the scattering cross section displayed
in Fig. 9, as discussed below. Taking into account the dispersion greatly 
modifies the numerical value of
$\alpha$, but other corrections are small in the case of positive
$\delta$.

In
the case of negative $\delta$, the interference pattern is strongly
modified.
This is illustrated in Fig. 7, where we plot the surface displacement for
$\delta = - 8$, $\beta = 5$ and
$\alpha = 0.5,\,1,\,1.5,\,2$. The spiral wave which is clearly seen for
$\alpha
= 2$ (Fig. 7-d) is very different from the corresponding figure of [I]
(see
Fig.
2-d of \cite{coumlu}). For larger values of $\beta$, shown in Fig. 8 for
which
$\beta = 10$, whith same values for $\delta$ and $\alpha$ as before, the
pictures are much more similar to the shallow water case.

This is confirmed by the plot of the absolute value of the correction to
the
Aharonov-Bohm scattering amplitude in both cases. The dashed line in Fig.
9
shows
this correction in the shallow water (non dispersive) case, and the solid
line
shows the same correction in the dispersive case, for a positive $\delta =
+ 8$.
Both corrections are almost the same, in agreement with the results of
Fig. 5
\& 6. The same is done for a negative $\delta = -8$ in Fig. 10. The
graphs,
this
time, differ enormously. In the dispersive case, the scattering is much
more
isotropic, and very different in amplitude. An obvious, but somewhat
formal, explanation of this
difference is the supplementary function $H^1_{m_-}(\beta \sqrt{|\delta |
-1})$
in \refpar{v20neg} compared to \refpar{v20pos}. Also, the index of this
function
is
$m_-(n)$ which takes imaginary values for small {\it positive} $n$, rather
than
for small {\it negative} $n$ as $m_+(n)$. This implies that the partial
amplitudes for
$\exp (-in\theta)$ and $\exp (in\theta)$ are much more similar to each other 
than in the
shallow water case, for which $m_-$ is absent, and also more similar than in the 
case of positive
$\delta$, where the decrease of the corresponding function is exponential.
The
appearance of an algebraically ($\propto 1/\sqrt{r}$) decreasing amplitude
associated to $m_-$ in the negative $\delta$ case restores the symmetry of
the
scattered wave.

A more physical explanation is as follows: Consider a plane wave
incident from the right on a
counterclockwise vortex, as in figures 5--8. Above the vortex, the wavefront
velocity is increased by
advection, whereas it is decreased below the vortex. Consequently the wave 
fronts should bend towards the bottom of the picture, as they do.
The other effect of the vortical flow is to add a wavelength below the
vortex, which means that the
wavenumber $k$ decreases below the vortex. Since $kh$ is assumed to be
small, for positive $\delta$ the
phase velocity increases with $k$. The phase velocity is thus smaller below
the vortex than above,
which enhances the bending of the wavefronts, and reinforces the strong
asymmetry in the interference
pattern of Fig. 5 \& 6, and in the scattering amplitude of Fig. 9. For
negative $\delta$, the
phase velocity {\it decreases} with $k$, and the effect of the dislocation
is to make the phase velocity
smaller for the part of the wavefront above the vortex. This effect
balances the effect of advection, and we
understand why the interference pattern (see Fig. 7 \& 8) and the scattering
amplitude (see Fig. 10)
are much more symmetric than in the positive $\delta$ case, or than in the
nondispersive ($\delta =
0$) case. It is also reasonable that this effect should be more important
for $\beta = 5$ than for
$\beta = 10$, because the relative variation in $k$ due to the dislocation
is greater in the former
case. The spiral waves are observed for negative $\delta$ because the
interference pattern almost
keeps rotational symmetry while smoothing the wavefront dislocation in the
forward direction.

As a last illustration, we compare the wave patterns predicted by the
shallow
water approximation, and by its first correction in powers of the fluid
depth, in
an experimentally accessible situation. We suppose the fluid to be pure
water,
of depth 1 mm.; the vortex radius is 1 cm., and the wavelength is 2 cm.
Thus
$kh = \pi/10$, and the approximation of the dispersion relation is
excellent.
The price to pay is the rather small value $\beta = \pi$. We take the
vortex
circulation such that $\alpha = 1$ in the shallow water approximation, for
which
$c \equiv \sqrt{gh} = 9.9 \; {\rm cm}/{\rm s}$. Taking the dispersion into
account, we get $\delta = 1.4$, $\cphase = 13.0 \;{\rm cm}/{\rm s}$,
$c_{\rm g} =
18.4 \;{\rm cm}/{\rm s}$ and
$\alpha = 0.41$. The difference in the respective numerical values of
$\alpha$
is the leading effect. The result is shown in Fig. 11. To obtain
quantitative
agreement with an experimental situation, it may be sufficient to keep the
shallow water approximation, but with the actual value of $\alpha$
obtained
in
the dispersive case \refpar{defalpha}. It should be interesting to use a
fluid
with a very small surface tension, in order to obtain a negative value of
$\delta$ while keeping a small value of $kh$, for which the wave pattern
should
be extremely different from the shallow fluid layer approximation.
All calculations were performed
using  Mathematica\cite{mathematica}.

\section{Concluding Remarks}
\label{sec:Conclusion}

We have computed the surface displacement due to a dispersive surface wave
interacting with a vertical vortex  when the vortex core
performs solid body rotation, the wavelength is small compared
to the vortex core radius and the particle velocities associated with the
wave are small compared to the particle velocities associated with the
vortex. When the parameter $\alpha = \nu \Gamma /2 \pi \cphase c_{\rm g}$
is of
order one or bigger, the wavefronts become dislocated. This parameter
depends,
in the dispersive case, both on the phase and group velocity of the wave,
and
tends smoothly toward the result of [I] in the nondispersive limit. We
thus
give
a proof, in a perturbative fashion, of the heuristic derivation of Berry
{\it et
al.}
\cite{berryetal}.

Formally, we proceed perturbatively around the shallow water limit, to obtain
a fourth order partial differential equation for the surface elevation 
associated with the surface wave. However, apart from some
technical
details, the solution is very similar to the nondispersive
case.
The
scattered waves interact strongly with the dislocated wavefronts and
produce
interference patterns. A dimensionless parameter $\delta$ quantifies the
relation between fluid
layer depth $h$ and capillary length $l_c$. When it is
positive
($h
< \sqrt{3} l_c$) the wave pattern is similar to the shallow water
case.
When it is negative, for large values of the circulation, the
wave
pattern is very different.

 We
hope that the calculations in the dispersive case will help the comparison
with experiments. Our calculations are valid when the approximation $\tanh
kh
\simeq kh - (kh)^3/3$ holds. This is a restrictive condition, but we
believe that, once dispersion is correctly taken into account in the
definition of
$\alpha$, the wave pattern given by the nondispersive case should be
roughly
similar to the observations. Some discrepancies are expected when the
fluid
depth is greater than the capillary length ($h
> \sqrt{3} l_c$), but in this case it is necessary to correctly
approximate
the
hyperbolic tangent by the first two terms of the series, and in practice
the
effect should be observable for a fluid of small surface tension (thus
small
capillary length) only.

\acknowledgements{ We than F. Melo for useful discussions and for pointing
out what turned out to be a serious flaw in an early version of this
paper. The work of F.L. is supported by Fondecyt
Grant 1960892 and a C\'atedra Presidencial en Ciencias. We gratefully
acknowledge a grant from ECOS-CONICYT.}

\newpage

\appendix
\section{Computational details}
\label{sec:Appendix}

In this Appendix, we explain how to calculate the coefficients in the
series
representations of Sec. \ref{sec:scattering}. We also discuss briefly
the convergence of the series involved in our solution
\refpar{fincoeur} and
\refpar{finextR}, since the sum in \refpar{finextAB} obviously converges
as
in
the shallow water case [I].

To get the solution of eqns. (\ref{rela}--\ref{reld}), we introduce the
column
vectors $V_x$ of the coefficients of the unknown $x_n$. They involve
multiple
derivatives of the Bessel functions, and may be simplified with the help
of
the following formulae \cite{grad}:
\begin{equation}
\cases{{\displaystyle Z'_p = {1\over 2}(Z_{p-1} - Z_{p+1})}, \cr
\quad \cr {\displaystyle I'_p = {1\over 2}(I_{p-1} + I_{p+1})}, \cr
\quad \cr {\displaystyle K'_p = -{1\over 2}(K_{p-1} + K_{p+1})},
\cr}\qquad
\cases{{\displaystyle Z_{p-1} = -Z_{p+1} + {2p\over z}Z_p,} \cr
\quad \cr {\displaystyle I_{p-1} = I_{p+1} + {2p\over z}I_p,} \cr \quad
\cr
{\displaystyle K_{p-1} = K_{p+1} - {2p\over z}K_p,} \cr}
\label{formBessel}
\end{equation}
where $z$ is the argument of the function, $p$ its index, and $Z$
represents either $J$ or $H^1$.

The column vectors thus read
\begin{equation}
V_a \equiv \pmatrix{1 \cr {\displaystyle n - \phi_n{J_{n+1}(\phi_n) \over
J_{n}(\phi_n)}} \cr {\displaystyle n(n-1) - \phi_n^2 +
\phi_n{J_{n+1}(\phi_n) \over J_{n}(\phi_n)}} \cr {\displaystyle
(n-1)[n(n-2) -
\phi_n^2] +
\phi_n[\phi_n^2 - (n^2 + 2)]{J_{n+1}(\phi_n)
\over J_{n}(\phi_n)}} },
\label{Veca}
\end{equation}
in the negative $\delta$ case, and the same expression with $\tilde \phi_n$ in the positive
$\delta$ case,
\begin{equation}
V_b \equiv \pmatrix{1 \cr {\displaystyle n + \varphi_n{I_{n+1}(\varphi_n)
\over
I_{n}(\varphi_n)}} \cr {\displaystyle n(n-1) + \varphi_n^2 -
\varphi_n{I_{n+1}(\varphi_n) \over I_{n}(\varphi_n)}} \cr {\displaystyle
(n-1)[n(n-2) +
\varphi_n^2] +
\varphi_n(\varphi_n^2 + n^2 + 2){I_{n+1}(\varphi_n)
\over I_{n}(\varphi_n)}} },
\label{Vecbpos}
\end{equation}
in the positive
$\delta$ case, or
\begin{equation}
V_b \equiv \pmatrix{1 \cr {\displaystyle n - \tilde{\varphi}_n{J_{n+1}(\tilde{\varphi}_n)
\over
J_{n}(\tilde{\varphi}_n)}} \cr {\displaystyle n(n-1) - \tilde{\varphi}_n^2 +
\tilde{\varphi}_n{J_{n+1}(\tilde{\varphi}_n) \over J_{n}(\tilde{\varphi}_n)}} \cr {\displaystyle
(n-1)[n(n-2) -
\tilde{\varphi}_n^2] +
\tilde{\varphi}_n[\tilde{\varphi}_n^2 - (n^2 + 2)]{J_{n+1}(\tilde{\varphi}_n)
\over J_{n}(\tilde{\varphi}_n)}} },
\label{Vecbneg}
\end{equation}
in the negative $\delta$ case,
\begin{equation}
V_d \equiv \pmatrix{-1 \cr {\displaystyle -m_+ + \beta{H^1_{m_+ +1}(\beta)
\over H^1_{m_+}(\beta)}} \cr {\displaystyle -m_+(m_+ -1) + \beta^2 -
\beta{H^1_{m_+ +1}(\beta) \over H^1_{m_+}(\beta)}} \cr {\displaystyle
(m_+ -1)[\beta^2 -m_+(m_+ -2) ] +
\beta( m_+^2 + 2-\beta^2 ){H^1_{m_+ +1}(\beta)
\over H^1_{m_+}(\beta)}} },
\label{Vecd}
\end{equation}
\begin{equation}
V_e \equiv \pmatrix{-1 \cr {\displaystyle -m_- + \beta \sqrt{1 +
\delta}\;{K_{m_-+1}\left(\beta \sqrt{1 + { \delta}}\right)
\over K_{m_-}\left(\beta \sqrt{1 + { \delta}}\right)}} \cr {\displaystyle
-m_-(m_- -1) -
\beta^2
\left(1 + {
\delta}\right) -
\beta \sqrt{1 + { \delta}}\;{K_{m_- +1}\left(\beta \sqrt{1 + {
\delta}}\right)
\over K_{m_-}\left(\beta \sqrt{1 + { \delta}}\right)}} \cr {\displaystyle
(1 -
m_- )\left[\beta^2
\left(1 + {
\delta}\right)+
m_-(m_- -2) \right] +}  \qquad \qquad\qquad\qquad\cr\qquad +
{\displaystyle
\beta
\sqrt{1 + {
\delta}}\left[ m_-^2 + 2 +
\beta^2
\left(1 + {
\delta}\right)
\right]{K_{m_-+1}\left(\beta \sqrt{1 + { \delta}}\right)
\over K_{m_-}\left(\beta \sqrt{1 + { \delta}}\right)}} },
\label{Vecepos}
\end{equation}
in the positive $\delta$ case, or
\begin{equation}
V_e \equiv \pmatrix{-1 \cr {\displaystyle -m_- + \beta
\sqrt{|\delta| - 1}\;{H^1_{m_-+1}\left(\beta \sqrt{|\delta | -1}\right)
\over H^1_{m_-}\left(\beta \sqrt{|\delta | -1}\right)}} \cr {\displaystyle
-m_-(m_- -1) +
\beta^2
\left(|\delta | - 1\right) -
\beta \sqrt{|\delta | -1}\;{H^1_{m_- +1}\left(\beta \sqrt{|\delta |
-1}\right)
\over H^1_{m_-}\left(\beta \sqrt{|\delta | -1}\right)}} \cr {\displaystyle
(
m_- - 1)\left[\beta^2
\left(|
\delta | - 1\right) -
m_-(m_- -2) \right] +}  \qquad \qquad\qquad\qquad\cr\qquad +
{\displaystyle
\beta
\sqrt{|\delta | -1}\left[ m_-^2 + 2 -
\beta^2
\left(|
\delta | - 1\right)
\right]{H^1_{m_-+1}\left(\beta \sqrt{|\delta | -1}\right)
\over H^1_{m_-}\left(\beta \sqrt{|\delta | -1}\right)}} },
\label{Veceneg}
\end{equation}
in the negative $\delta$ case. The column vector $V_c$ for the
coefficients
of
$c_n$ in the right hand member of (\ref{rela}--\ref{reld}) reads
\begin{equation}
{V_c \over (-i)^{m_+}} \equiv \pmatrix{J_{m_+}(\beta) \cr {\displaystyle
m_+  J_{m_+}(\beta) -
\beta{J_{m_+ +1}(\beta)}} \cr {\displaystyle [m_+(m_+ -1) -
\beta^2]J_{m_+}(\beta) +
\beta{J_{m_+ +1}(\beta)} } \cr {\displaystyle
(m_+ -1)[ m_+(m_+ -2) - \beta^2]J_{m_+}(\beta) +}\qquad\qquad\qquad \cr
\qquad\qquad\qquad\qquad {\displaystyle +
\beta[\beta^2-( m_+^2 + 2)]{J_{m_+ +1}(\beta)} }}.
\label{Vecc}
\end{equation}
Now, formally, the coefficient $a_n$ (say) is expressed as the
ratio of two determinants,
\begin{equation}
a_n = {\left|V_cV_bV_dV_e\right| \over \left|V_aV_bV_dV_e\right|},
\label{anformel}
\end{equation}
and is a function of $n$, $\alpha$, $\beta$ and $\delta$.

Let us discuss briefly the asymptotic behavior of those coefficients. For
large
$n$, $m_+ \sim n$, and we can safely assume that $Z_{n+1}/Z_n = O(1)$, in
the
case of a constant argument, $\beta$, $\beta\sqrt{1 +\delta}$,
$\beta\sqrt{|\delta| - 1}$, or in the case of an argument of order $n$,
such
as
$\phi_n$ or
$\varphi_n$. This is demonstrated for ordinary Bessel functions in [I],
and
may
be deduced in the same fashion for hyperbolic Bessel functions, which have
similar behaviors for large values of the index \cite{abrastegun}. From
the
calculations of ref. [I], we easily get the dominent behavior for large
$n$ of the vectors $V_x$, and thus of the coefficients $x_n$. If $x_n$ is
not
equal to $c_n$,
\begin{equation}
V_x \sim \pmatrix{1 \cr n \cr n^2 \cr n^3}
\label{asymptVxneqc}
\end{equation}
so that the determinant in the denominator of any expression such as
\refpar{anformel} behaves like
\begin{equation}
\left|V_aV_bV_dV_e\right| \sim n^6.
\label{asymptdetden}
\end{equation}
We also have that
\begin{equation}
V_c \sim \pmatrix{1 \cr n \cr n^2 \cr n^3}J_n(\beta) \sim \pmatrix{1 \cr n
\cr
n^2 \cr n^3}{1\over \sqrt{n}}\left({e\over n}\right)^n,
\label{asymptVcneqc}
\end{equation}
so that the determinant in the numerator of \refpar{anformel} behaves like
\begin{equation}
\left|V_cV_bV_dV_e\right| \sim n^{5/2}\left({e\over n}\right)^n,
\label{asymptdetnum}
\end{equation}
and the coefficients $x_n$ decrease sufficiently quickly to ensure the
convergence of the series in \refpar{fincoeur} and
\refpar{finextR}. The convergence is uniform for $r<a$, because the
support
of
the functions is compact, but not for $r>a$.


\newpage
\begin{figure}
\caption{Absolute magnitude of coefficients $a_n$ (a) and $c_
n$ (b) versus $n$, in a log-linear plot, for a positive $\delta = +8$, and for 
values of dislocation parameter $\alpha$ and dimensionless radius $\beta$ 
$(\alpha, \beta)
=(0.5,10)$,
denoted by dots,
$(\alpha,
\beta) =(1.5,10)$, denoted by empty circles $\circ$, $(\alpha, \beta)
=(0.5,5)$,
denoted by filled circles $\bullet$
 and $(\alpha, \beta) =(1.5,5)$, denoted by empty squares $\Box$. Note the
asymmetry with respect to $n \rightarrow -n$ }
\label{figure-1}
\end{figure}

\begin{figure}
\caption{Same as figure 1, for the coefficients $d_n$ (a) and $e_
n$ (b).}
\label{figure-2}
\end{figure}

\begin{figure}
\caption{Same as figure 1, for a negative $\delta = -8$. }
\label{figure-3}
\end{figure}

\begin{figure}
\caption{Same as figure 2, for a negative $\delta = -8$.}
\label{figure-4}
\end{figure}

\begin{figure}
\caption{Density plot of the surface elevation for the total wave patterns
for $\delta = +8$ and $\beta = 5$, for several values of $\alpha$.
Respectively $\alpha = 0.5$  : (a),
$\alpha = 1$ : (b),
$\alpha = 1.5$ : (c),
$\alpha = 2$ : (d). The greyscale is linear with surface  amplitude
(arbitrary
units). The dark ring  indicates the vortex
location, and the vortex rotates counterclockwise. The wave is incident
from the right. The box size is $10\times 10$ in
units
of $a$, the vortex radius.}
\label{figure-5}
\end{figure}

\begin{figure}
\caption{Same as figure 5, for $\delta = +8$, $\beta = 10$, and several
values
of $\alpha$. Respectively $\alpha = 0.5$ : (a), $\alpha = 1$ :
(b),
$\alpha = 1.5$ :  (c), $\alpha = 2$ : (d).}
\label{figure-6}
\end{figure}

\begin{figure}
\caption{Same as figure 5, for $\delta = -8$, $\beta = 5$, and several
values
of $\alpha$. Respectively $\alpha = 0.5$ : (a), $\alpha = 1$ :
(b),
$\alpha = 1.5$ :  (c), $\alpha = 2$ : (d).}
\label{figure-7}
\end{figure}

\begin{figure}
\caption{Same as figure 5, for $\delta = -8$, $\beta = 10$, and several
values
of $\alpha$. Respectively $\alpha = 0.5$ : (a), $\alpha = 1$ :
(b),
$\alpha = 1.5$ :  (c), $\alpha = 2$ : (d).}
\label{figure-8}
\end{figure}

\begin{figure}
\caption{Polar plot of the absolute value
of the correction to the Aharonov-Bohm (i.e. point)
scattering amplitude, in the case of non dispersive waves (dashed line)
and
in the case of dispersive waves (solid line), in the case of a positive
$\delta
= +8$, for
$\beta = 5$ and
$\alpha = 0.25$ : (a),
$\alpha = 0.5$ : (b),
$\alpha = 1.0$ : (c), $\alpha = 1.25$ : (d). The vortex location is marked
by the
large dot; the vortex rotates counterclockwise.  }
\label{figure-9}
\end{figure}

\begin{figure}
\caption{Same as figure 9, but for a negative $\delta = -8$, for $\beta =
5$ and
$\alpha = 0.25$ : (a),
$\alpha = 0.5$ : (b),
$\alpha = 1.0$ : (c), $\alpha = 1.25$ : (d). The vortex location is marked
by the
large dot; the vortex rotates counterclockwise.  In this case the dispersive 
wave scatters very differently from the nondispersive wave. }
\label{figure-10}
\end{figure}

\begin{figure}
\caption{Density plot of the surface elevation for the total wave pattern
calculated in the shallow water approximation, (a), and to first order in
fluids
depth (b). The greyscale is linear with surface  amplitude (arbitrary
units). The dark ring  indicates the vortex
location, and the vortex rotates counterclockwise. The incident
wave
comes from the right. The box size is $20\times 20$ in
units
of $a$, the vortex radius. $\beta = \pi$ in both cases, but
$\alpha =
1$ in the shallow water case (a), and $\alpha = 0.41$, $\delta = 1.4$ in
the
dispersive case (b).}
\label{figure-11}
\end{figure}


\clearpage
\newpage

\begin{table}
\begin{center}
\begin{tabular}{|c|c|c|c|c|}
\multicolumn{1}{|l|}{} & \multicolumn{3}{c|}{C-G waves} &
S-W waves \\ \hline
$\lambda \,({\rm cm})$ & 0.1 & 0.5 & 1 & 2 \\ \hline
$\cphase \,({\rm cm}/{\rm s})$  & 68 & 32 &25 &  30 \\ \hline
$T^{\rm GW}\,({\rm s})$  & $10^{-7}$ & 0.0018 & 0.08 & 0.6 \\ \hline
$T^{\rm CW}\,({\rm s})$ &  0.013 & 0.3  & 1.26  & -- \\ \hline
Wave period $($s$)$  & 0.0015 & 0.0156 &0.04  & 0.07 \\ 
\end{tabular}
\end{center}
\caption{Attenuation times for capillary-gravity (C-G)
waves and shallow water (S-W) waves, compared to the period of the wave.
The
attenuation times are defined by \refpar{attentime}
\label{tab:OGattenuation}}

\end{table}


\clearpage
\newpage

\begin{figure}
\vskip -4truecm
\epsfysize=200mm
\centerline{\epsfbox{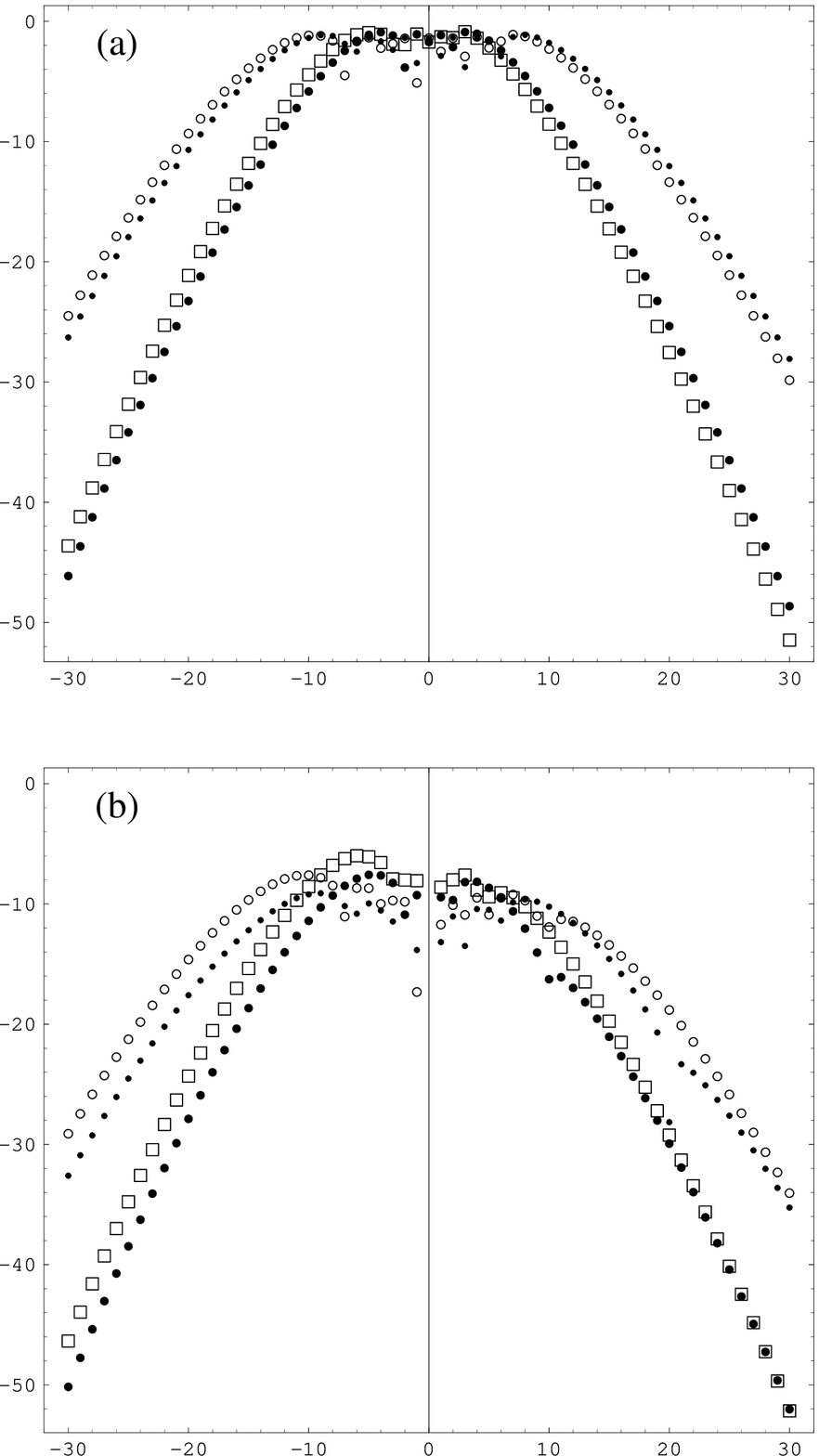}}
\vskip 2truecm
C. Coste {\it et al.}, Figure \ref{figure-1}
\end{figure}

\newpage

\begin{figure}
\vskip -3truecm
\epsfysize=200mm
\centerline{\epsfbox{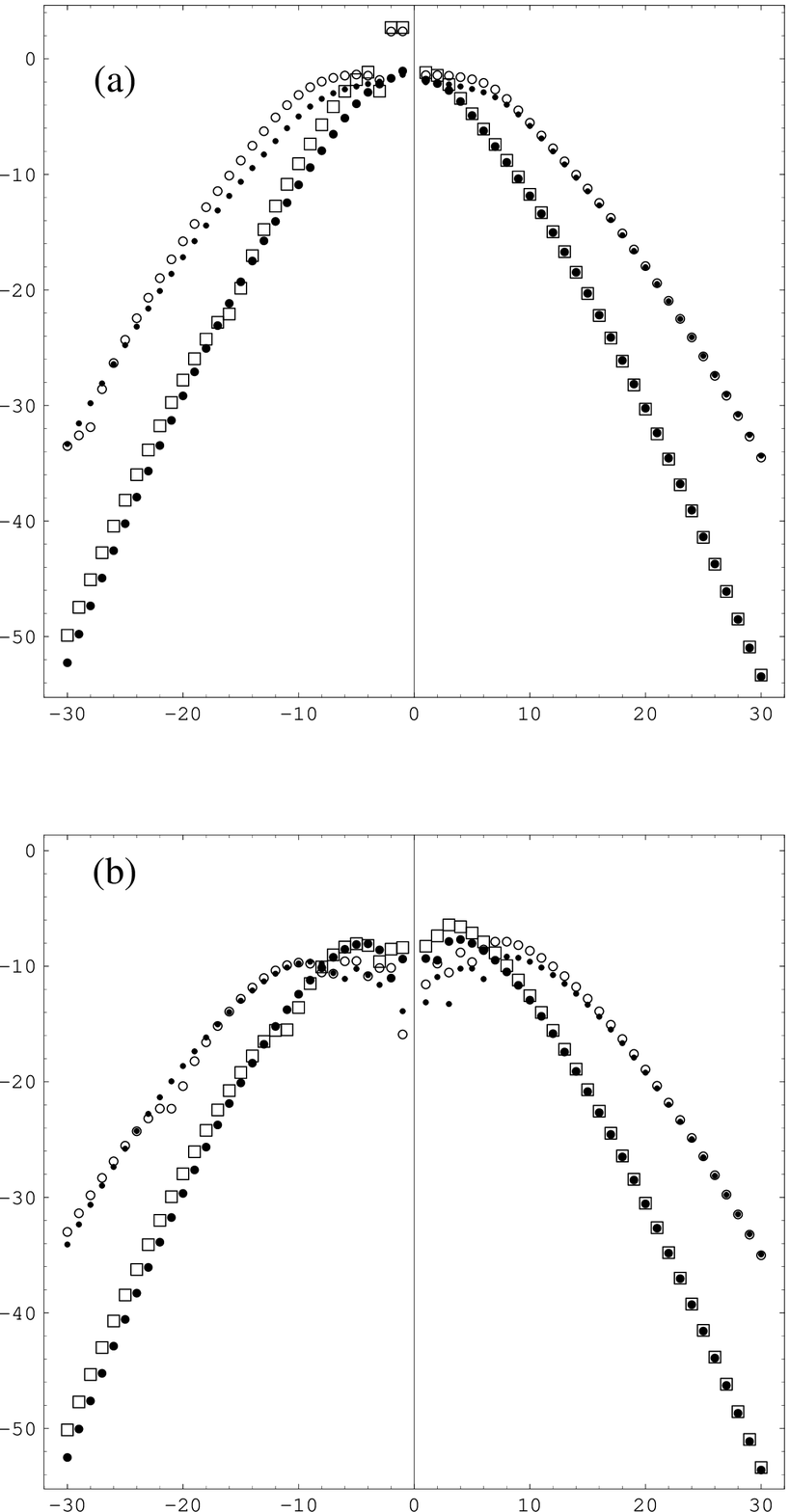}}
\vskip 2truecm
C. Coste {\it et al.}, Figure \ref{figure-2}
\end{figure}

\newpage

\begin{figure}
\vskip -3truecm
\epsfysize=219mm
\centerline{\epsfbox{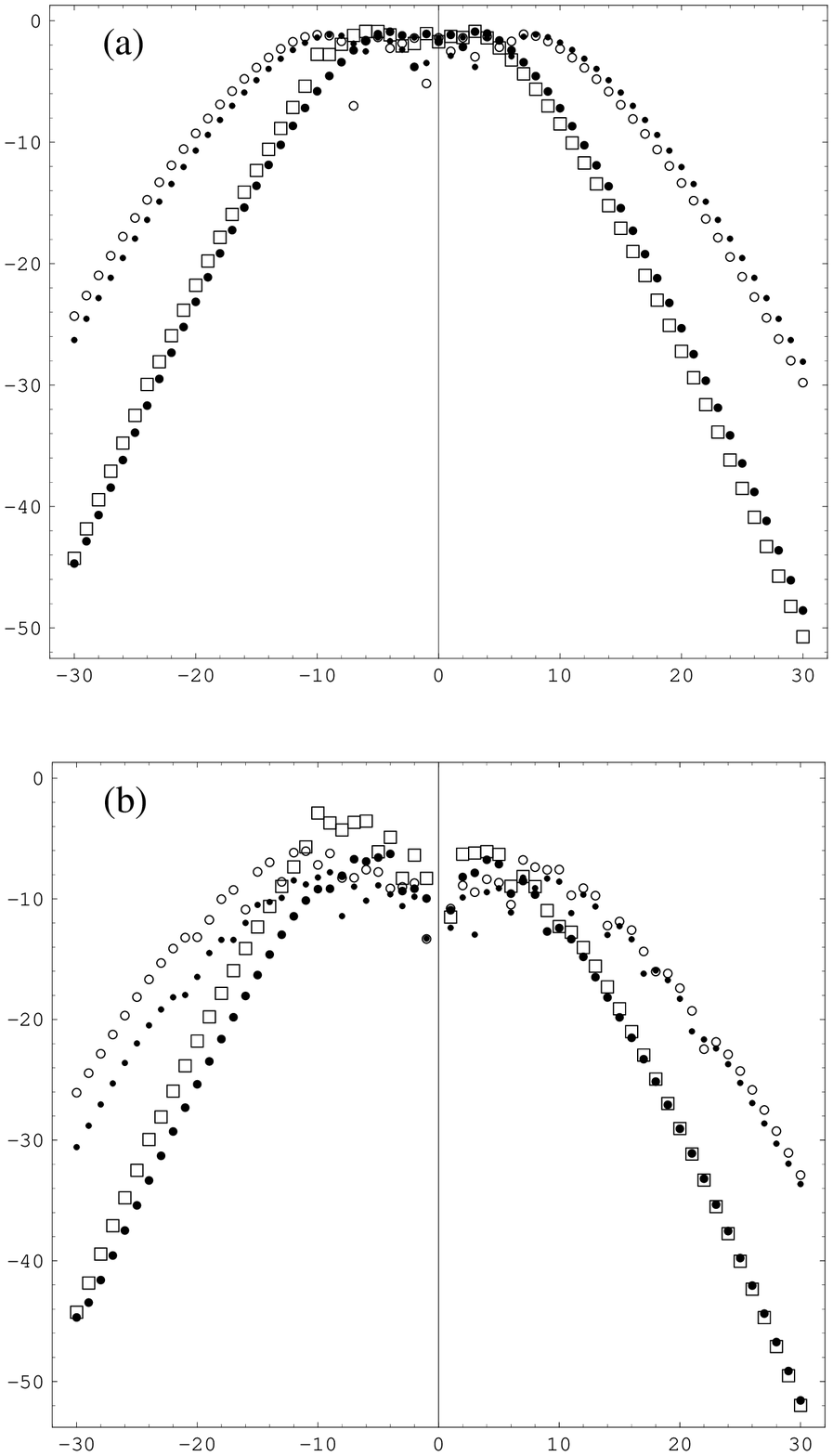}}
\vskip 2truecm
C. Coste {\it et al.}, Figure \ref{figure-3}
\end{figure}

\newpage

\begin{figure}
\vskip -5truecm
\epsfysize=200mm
\centerline{\epsfbox{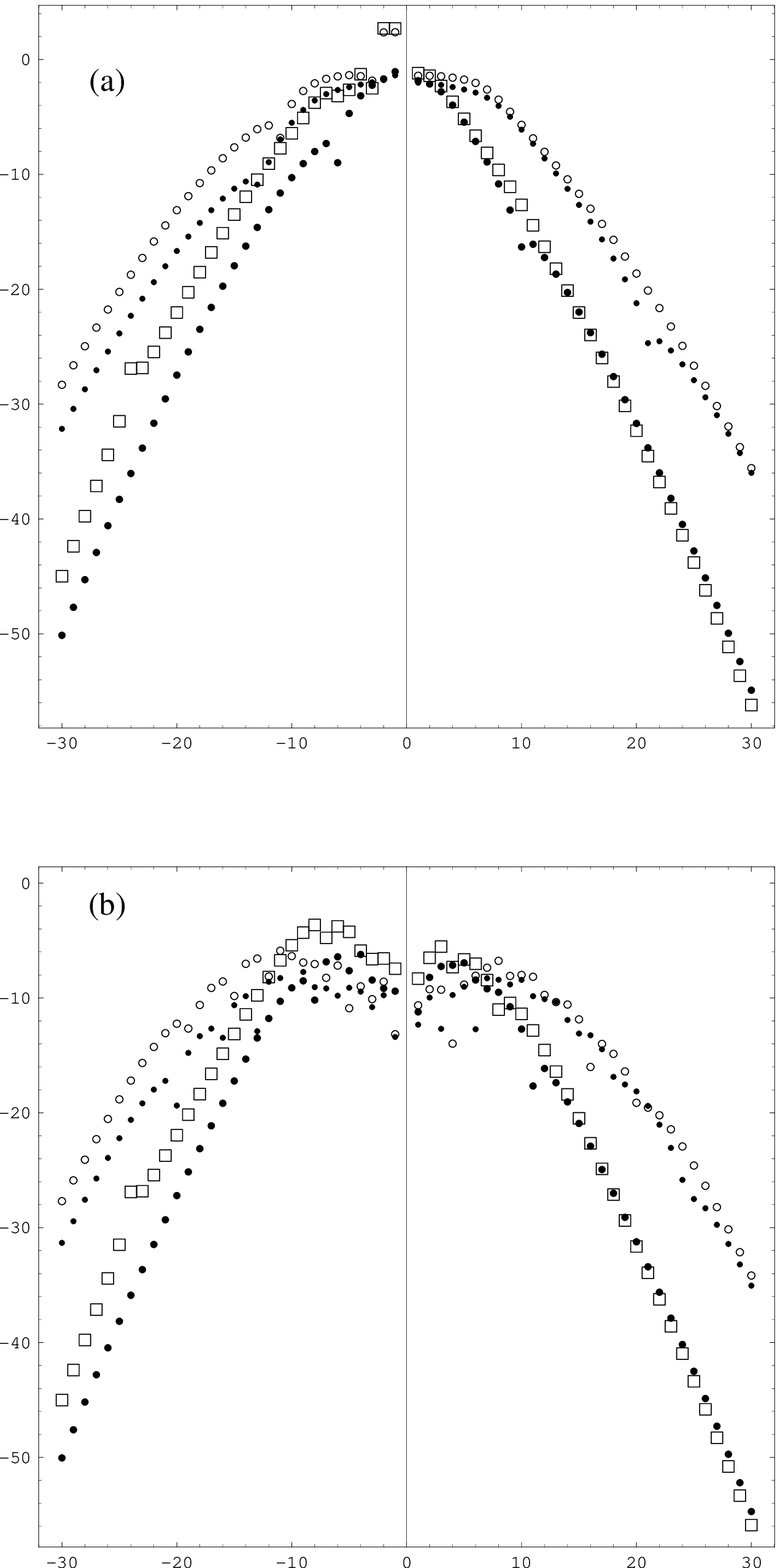}}
\vskip 2truecm
C. Coste {\it et al.}, Figure \ref{figure-4}
\end{figure}

\newpage

\begin{figure}
\centerline{\epsfbox{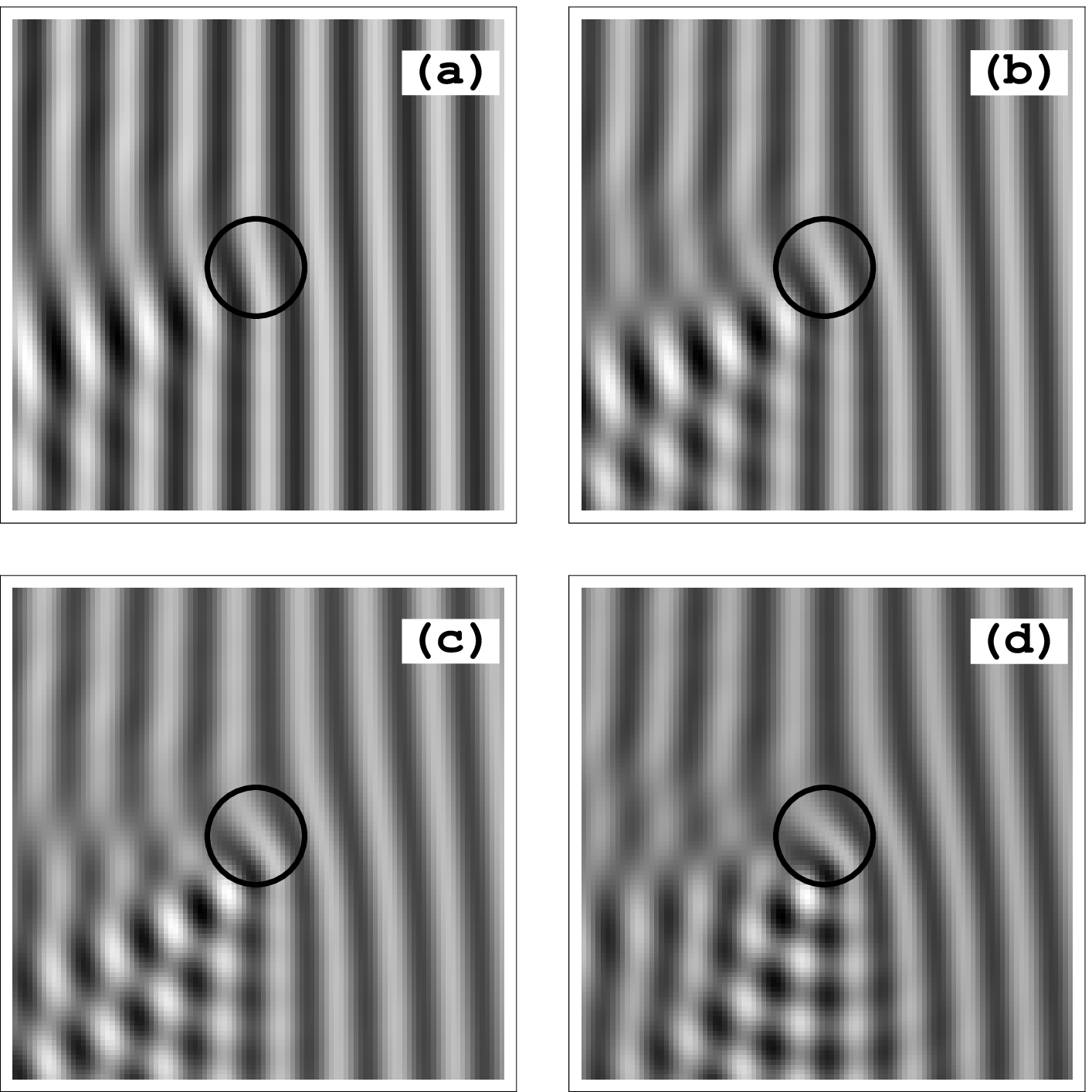}}
\vskip 3truecm
C. Coste {\it et al.}, Figure \ref{figure-5}
\end{figure}

\newpage

\begin{figure}
\centerline{\epsfbox{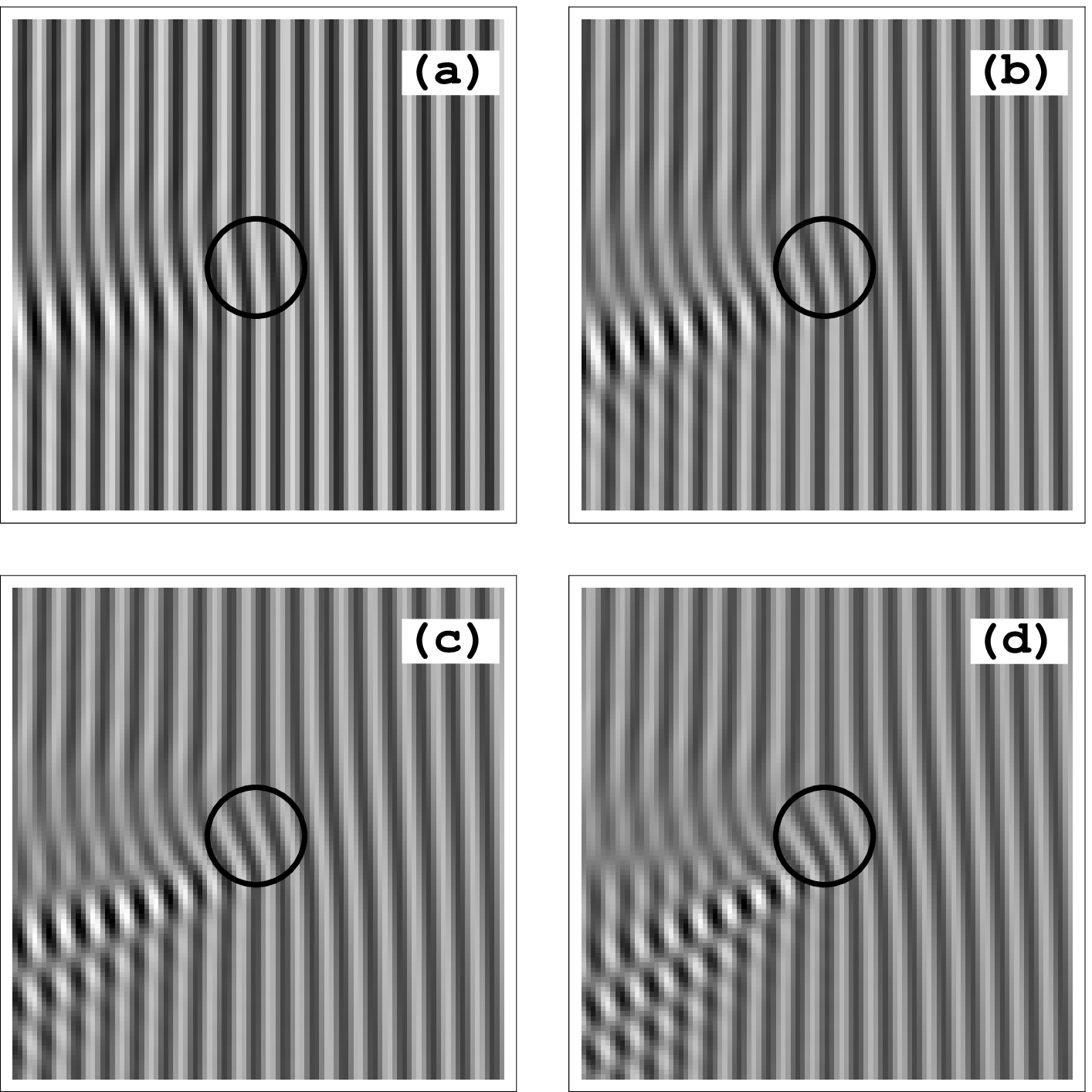}}
\vskip 3truecm
C. Coste {\it et al.}, Figure \ref{figure-6}
\end{figure}

\newpage

\begin{figure}
\centerline{\epsfbox{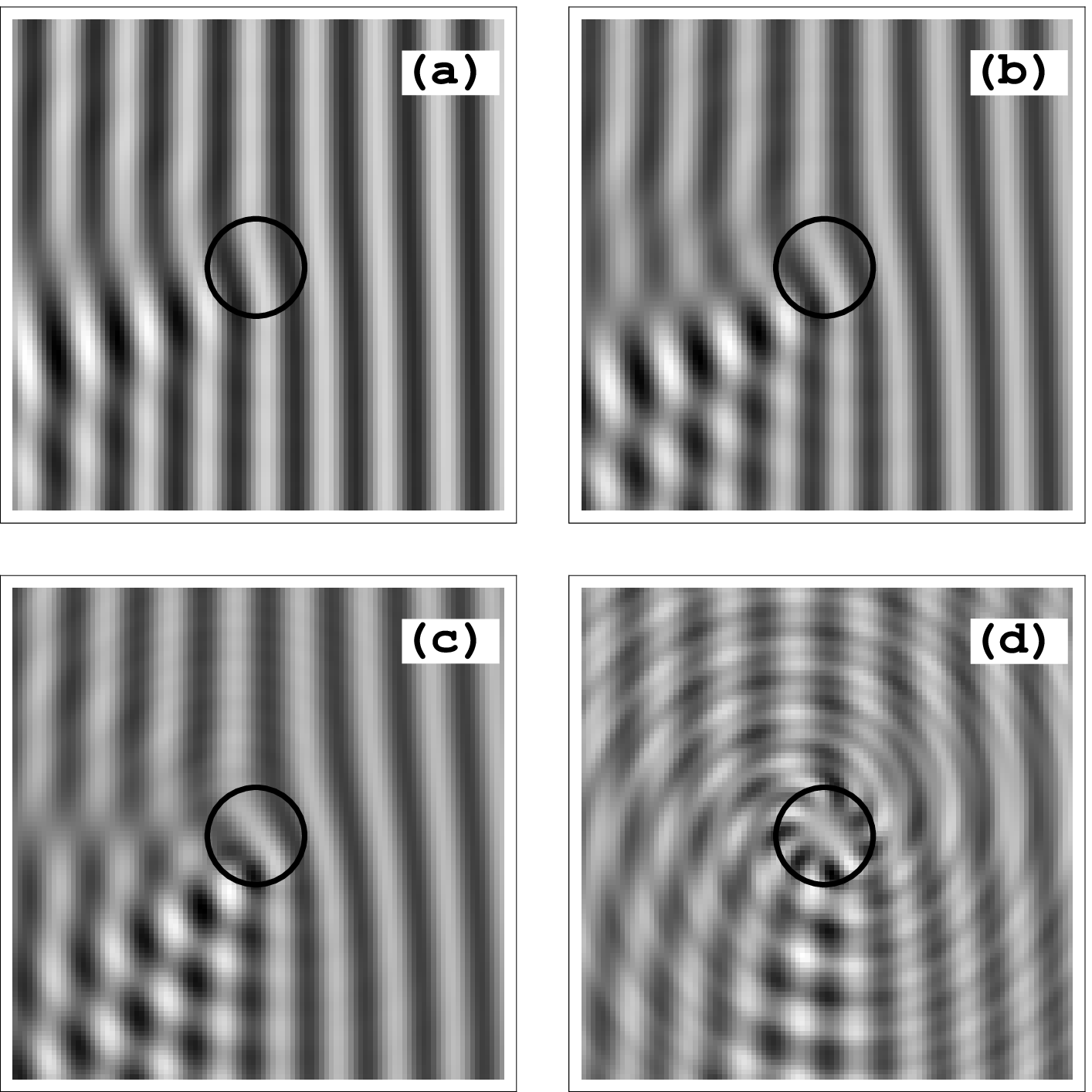}}
\vskip 3truecm
C. Coste {\it et al.}, Figure \ref{figure-7}
\end{figure}

\newpage

\begin{figure}
\centerline{\epsfbox{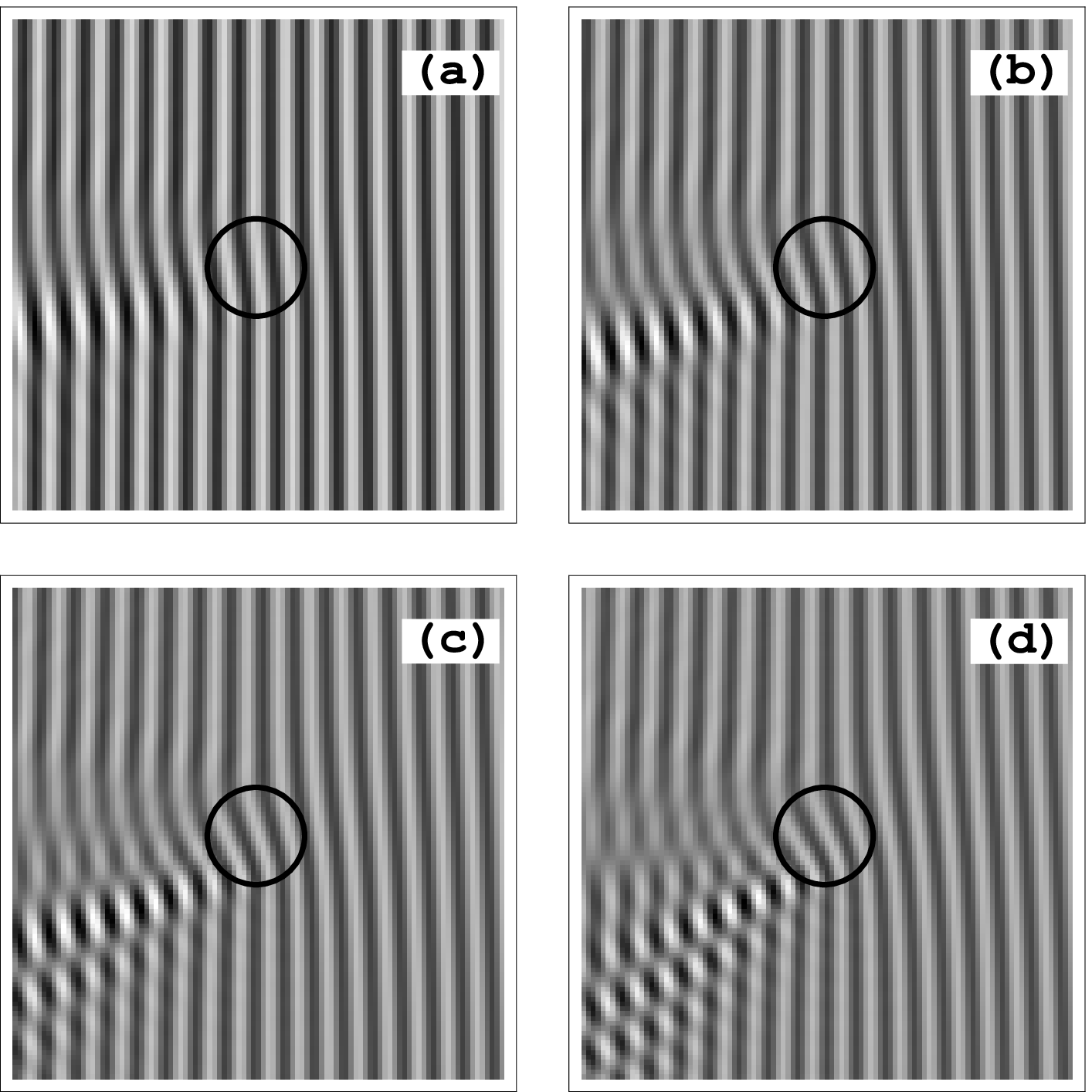}}
\vskip 3truecm
C. Coste {\it et al.}, Figure \ref{figure-8}
\end{figure}

\newpage

\begin{figure}
\vskip -5truecm
\epsfysize=130mm
\centerline{\epsfbox{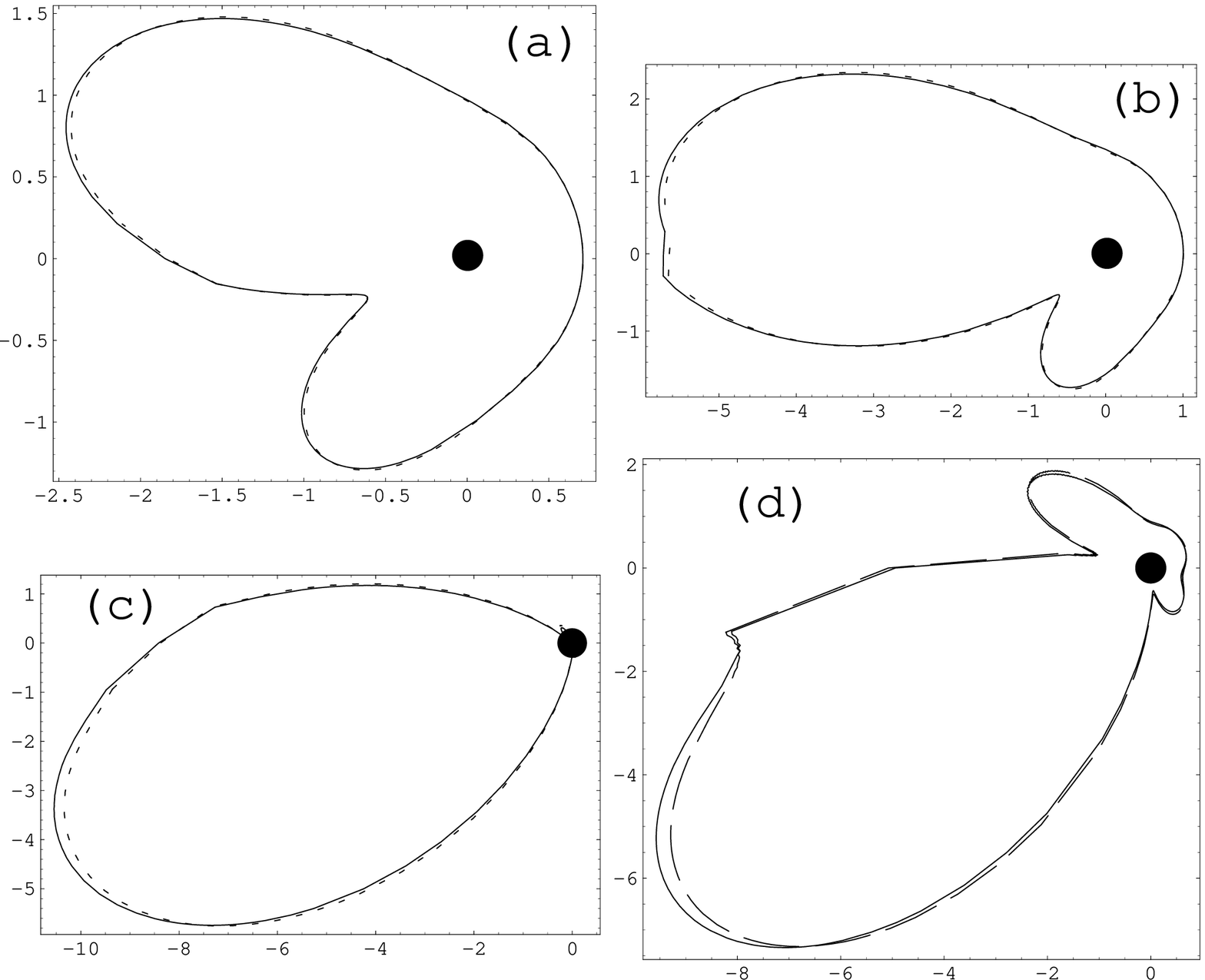}}
\vskip 2truecm
C. Coste {\it et al.}, Figure \ref{figure-9}
\end{figure}

\newpage

\begin{figure}
\vskip -5truecm
\epsfysize=150mm
\centerline{\epsfbox{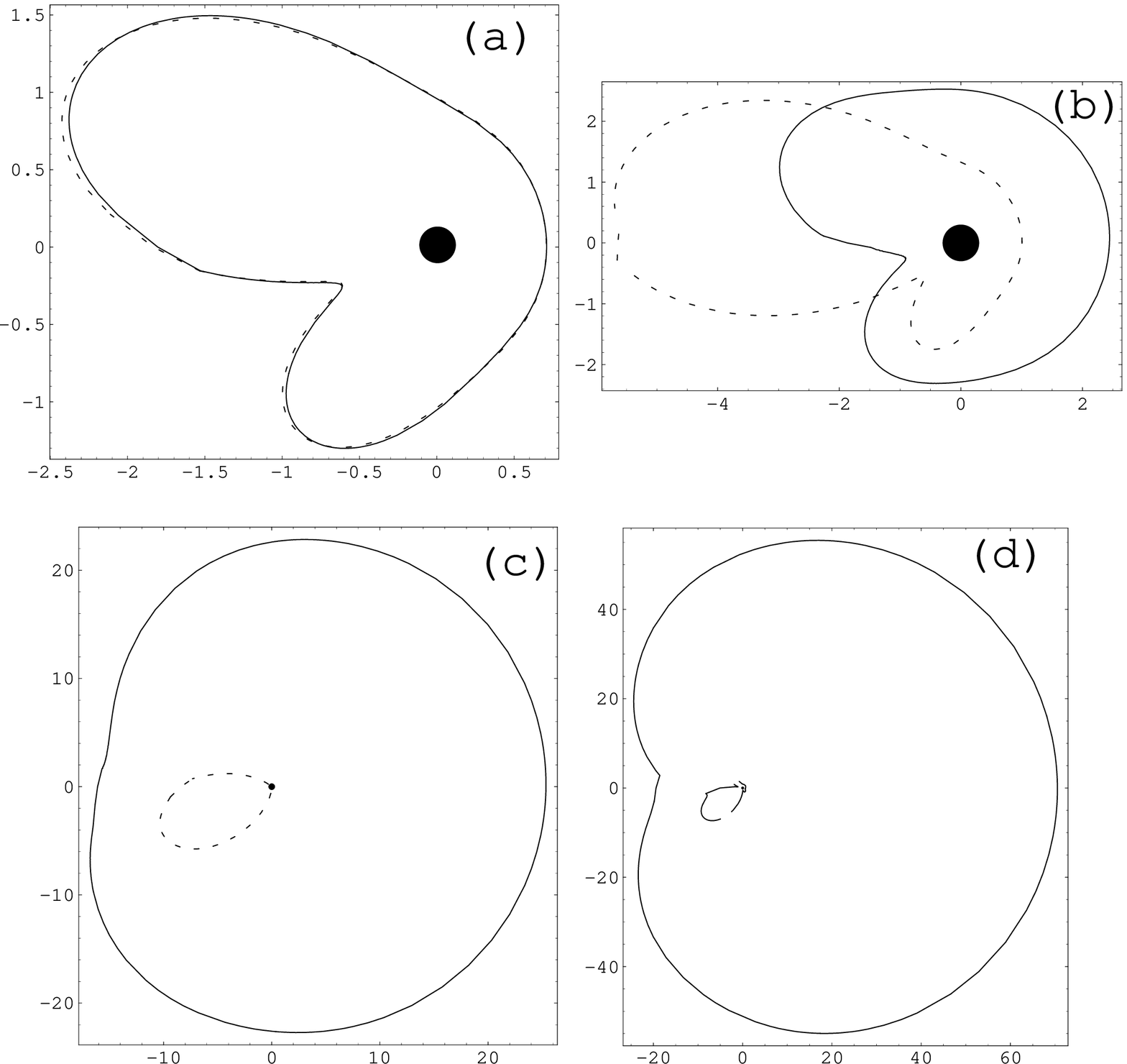}}
\vskip 2truecm
C. Coste {\it et al.}, Figure \ref{figure-10}
\end{figure}

\newpage

\begin{figure}
\vskip -2truecm
\epsfysize=80mm
\centerline{\epsfbox{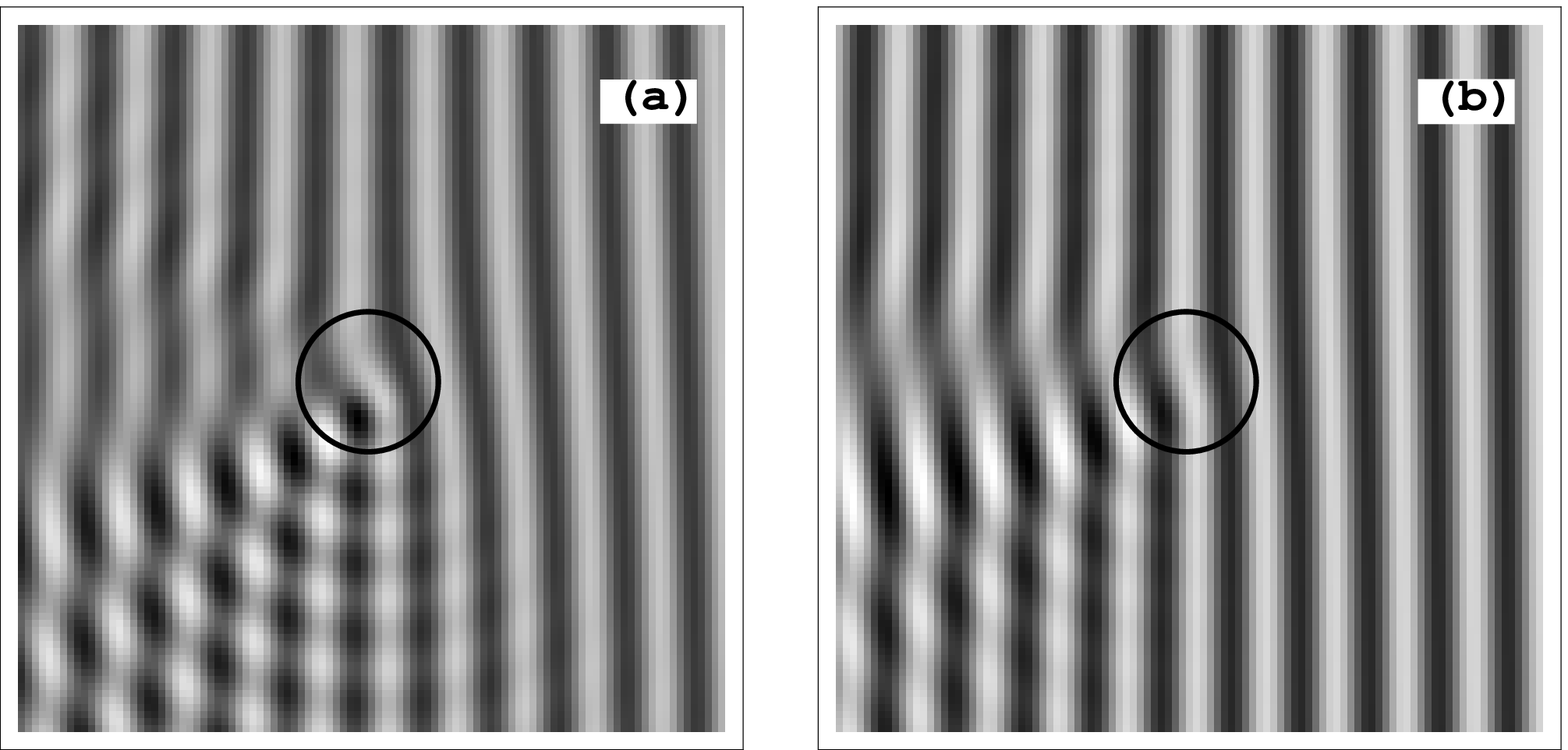}}
\vskip 5truecm
C. Coste {\it et al.}, Figure \ref{figure-11}
\end{figure}


\begin{references}


\bibitem{coumlu} C. Coste, M. Umeki and F. Lund, ``Scattering of
dislocated wavefronts by vertical vorticity and the
Aharonov-Bohm effect I: Shallow water'', Preprint, 1998, preceding
paper.

\bibitem{berryetal} M. V. Berry, R. G. Chambers, M. D. Large, C. Upstill
and J. C. Walmsley, {\it Eur. J. Phys.} {\bf 1}, 154 (1980).

\bibitem{umlu} M. Umeki and F. Lund, {\it Flu. Dyn. Res.} {\bf 21}, 201
(1997).

\bibitem{vivancomelo}  F. Vivanco and F. Melo, {\it Preprint}   (1998).

\bibitem{landau6}  L. D. Landau and E. M. Lifshitz, {\it Fluid Mechanics},
 2nd Ed., Pergamon (1987).

\bibitem{note1} It is tempting to take $z = h$ in \refpar{bcone1}
and \refpar{bcone2}; this is obviously consistent for linear water waves,
for
which $\eta$ is a small quantity, but here the surface elevation takes
into
account the vortical flow as well. The justification of this approximation
requires
an estimate for $\eta_0$. This is done in the discussion that follows
Eqns. \refpar{approxbcone1} and \refpar{approxbcone2}.


\bibitem{foot1new} In this case, the restriction $u \ll U$ imposed in
Section
II will break down when $r$ is very small or very large. At those
points, however, the condition $u \ll c$, implicit in the derivation of
Eqn. (\ref{eqfin}), assures that nonlinear terms can still be neglected.

\bibitem{morseingard}  P.M. Morse and K.U. Ingard, {\it Theoretical
acoustics},
 1st Ed., Princeton University Press (1986).

\bibitem{mathematica} S. Wolfram, {\it The Mathematica Book}, Third
Edition,
Cambridge University Press (1996).

\bibitem{grad} I. S. Gradshteyn and I. M. Ryshik, {\it Table of
Integrals, Series,  and Products}, Academic, 1980.

\bibitem{abrastegun} M. Abramowitz and I.A. Stegun, {\it Handbook of
mathematical functions}, Dover, 1972.

\end{references}
\end{document}